\documentclass[aps,prl,reprint, superscriptaddress]{revtex4-1}
\usepackage{blindtext}
\def \beq{\begin{eqnarray}}
\def \eeq{\end{eqnarray}}
\def \del{\partial}
\newcommand{\mbf}[1]{\boldsymbol{#1}}
\usepackage{amsmath}
\usepackage{amssymb}
\usepackage{graphicx}
\usepackage{graphicx,xcolor}
\usepackage{enumitem}
\newcommand{\rev}[1]{{\color{black}#1}}
\usepackage{float}
\usepackage[colorlinks=true,linkcolor=blue,citecolor=blue,urlcolor=blue]{hyperref}

\begin{document}

\title{Emergent polar order in non-polar mixtures with non-reciprocal interactions}

\author{Giulia Pisegna}
\affiliation{Max Planck Institute for Dynamics and Self-Organization (MPIDS), D-37077 Göttingen, Germany}
\author{Suropriya Saha}
\affiliation{Max Planck Institute for Dynamics and Self-Organization (MPIDS), D-37077 Göttingen, Germany}
\author{Ramin Golestanian}
\email{ramin.golestanian@ds.mpg.de}
\affiliation{Max Planck Institute for Dynamics and Self-Organization (MPIDS), D-37077 Göttingen, Germany}
\affiliation{Rudolf Peierls Centre for Theoretical Physics, University of Oxford, Oxford OX1 3PU, United Kingdom}

\begin{abstract}
Phenomenological rules that govern the collective behaviour of complex physical systems are powerful tools because they can make concrete predictions about their universality class based on generic considerations, such as symmetries, conservation laws, and dimensionality. While in most cases such considerations are manifestly ingrained in the constituents, novel phenomenology can emerge when composite units associated with emergent symmetries dominate the behaviour of the system. We study a generic class of active matter systems with non-reciprocal interactions and demonstrate the existence of true long-range polar order in two dimensions and above, both at the linear level and by including all relevant nonlinearities in the Renormalization Group sense. We achieve this by uncovering a mapping of our scalar active mixture theory to the Toner-Tu theory of dry polar active matter by employing a suitably defined polar order parameter. We then demonstrate that the complete effective field theory -- which includes all the soft modes and the relevant nonlinear terms -- belongs to the (Burgers-) Kardar-Parisi-Zhang universality class. This classification allows us to prove the stability of the emergent polar long-range order in scalar non-reciprocal mixtures in two dimensions, and hence a conclusive violation of the Mermin-Wagner theorem.
\end{abstract}

\maketitle

The field of active matter describes the collective behaviour of non-equilibrium systems, which are composed of units that break detailed-balance at the smallest scale \cite{gompper2020,Prost2015,toner1998flocks}, and are often classified based on the symmetries of these microscopic units \cite{marchetti2013hydrodynamics}. It is possible, however, that spontaneously formed composite units can lead to the emergence of physical behaviour that is completely different from what is expected for the system. Examples of such occurrences in condensed matter physics, which can often -- though not always -- accompany emergent symmetries, include formation of Cooper-pairs in the BCS theory of superconductivity \cite{BCS1957} as well as fractionalization and spin-charge separation in models of high-temperature superconductivity \cite{LeeRMP2006}. In active matter, a rare feature presents itself where non-reciprocal interactions (or action-reaction symmetry breaking) can lead to the emergence of polarity in non-polar mixtures \cite{soto2014self}, as afforded at small scales by the physics of phoretic active matter \cite{Golestanian2019phoretic}, and observed in experiments \cite{niu2018dynamics,meredith2020predator}.


A fundamentally important feature of non-reciprocal interactions -- when Newton's third law is apparently violated because of mutually asymmetric response in systems out of equilibrium \cite{ivlev2015statistical} -- is its inherent connection with time-reversal symmetry breaking, which is a notion commonly associated with self-propulsion in the context of active matter \cite{3SS}. This feature has been studied widely in polar \cite{uchida2010synchronization,Saha2019,Dadhichi_PhysRevE.101.052601,fruchart2021non, kreienkamp2022clustering,Loos2023}
and scalar \cite{agudo2019active,saha2020scalar,you2020nonreciprocity,frohoff2021suppression,dinelli2023non} 
active matter, alongside other types of phenomenology including the ability to sustain novel spatio-temporal patterns \cite{uchida2010synchronization,saha2020scalar,you2020nonreciprocity,fruchart2021non, weis2022coalescence, saha2022effervescent}, spontaneous chiral symmetry breaking \cite{fruchart2021non}, capability to design shape-shifting self-organizing structures \cite{soto2015self,osat2023non}, and proposals for fast and efficient self-organization of primitive metabolic cycles at the origin of life \cite{OuazanReboul2023-I,OuazanReboul2023-II,OuazanReboul2023-III}.

Here, we formally investigate the occurrence of {\em emergent polar symmetry} due to chasing interactions in non-reciprocal mixtures, in the context of the recently introduced non-reciprocal Cahn-Hilliard (NRCH) model \cite{saha2020scalar, you2020nonreciprocity}. Using a suitably defined polar order parameter \rev{(introduced previously in Ref. \cite{saha2020scalar})}, which measures the {\em coherence} between the two species, we derive \rev{a novel}  effective governing dynamics \rev{for} the emergent polar order field, and explore its connections with the conventional theories of polar flocks \cite{toner1998flocks,toner1995long,Toner2012PRL,Toner2012PRE}. \rev{The new framework enables us to formally} investigate the possibility of polar ordering, as well as its stability to fluctuations. \rev{This is achieved} both at the linear level using a comprehensive coarse-graining of the microscopic description of the system, as well as the fully nonlinear description that exploits a mapping to the Kardar-Parisi-Zhang (KPZ) universality class \cite{kardar1986dynamic,forster1977large,Frey1994}. Both linear and nonlinear descriptions exhibit true long-range polar order in the system in any dimension higher than one, in violation of the Mermin-Wagner theorem \cite{mermin1966absence}.

\subsection*{The model} 
\begin{figure*}[t]
\centering
\includegraphics[width=0.8\linewidth]{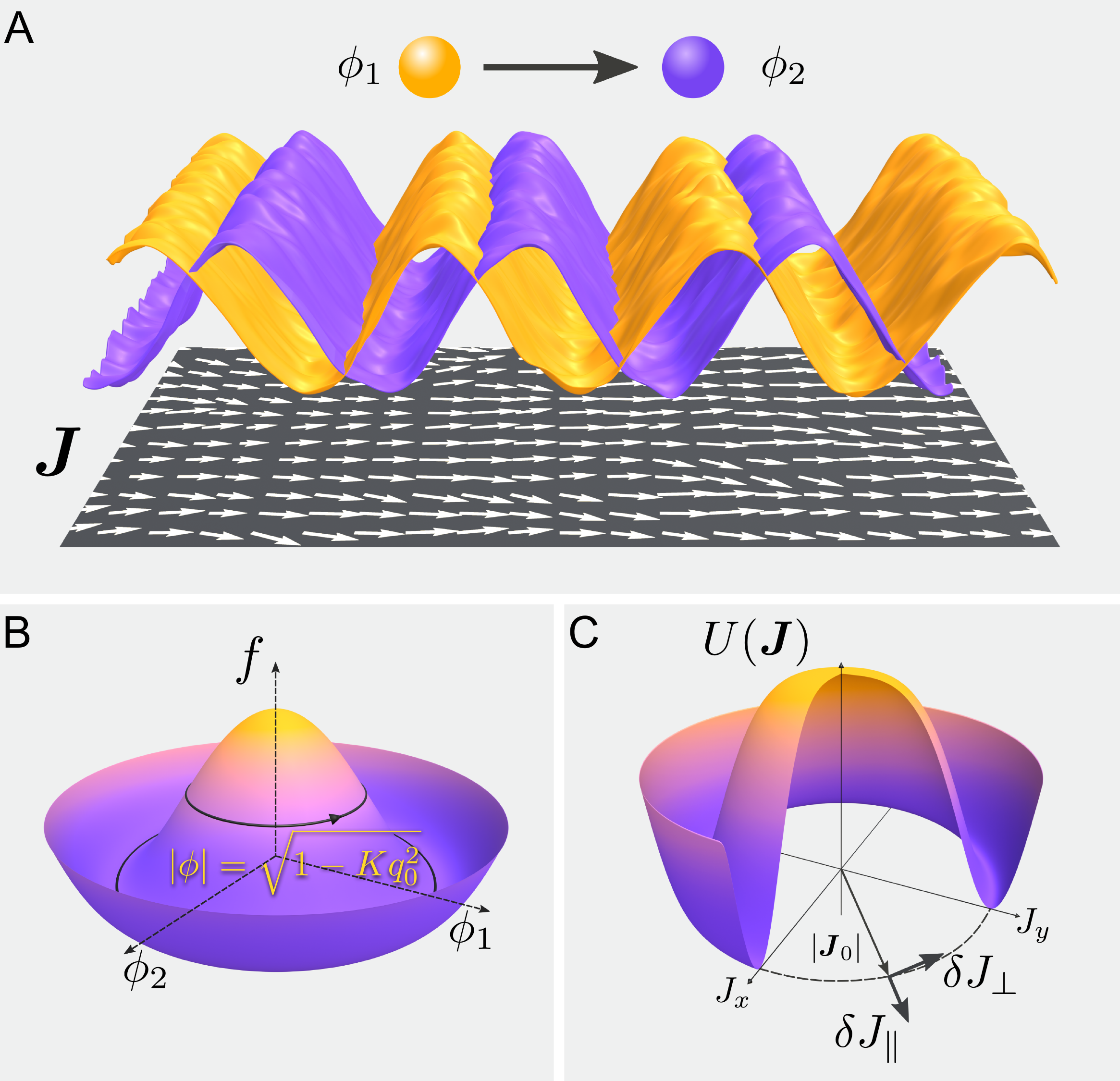}
\caption{{Emergent polar order in active scalar mixtures.}
{\em (A)} 
Due to non-reciprocal interactions, type-1 particles chase type-2 particles, while both being governed by conserved dynamics. The corresponding concentrations $\phi_1$ and $\phi_2$ exhibit phase-shifted spatio-temporal oscillations that we map onto the active polar order parameter $\mbf J$. The corresponding configuration for the polar field $\mbf J$ 
is shown in the plane below the waves. 
The vector fields and the waves are evaluated from numerical simulations of Eq. \ref{phi1} (see {\em Materials and Methods}). 
{\em (B)} 
Representation of the free-energy density in the $(\phi_1,\phi_2)$ space. The wave state with amplitude $|\phi|=\sqrt{1-K q_0^2}$ corresponds to an energetic level higher than the bottom of the potential well. 
{\em (C)} 
The ordered state $\mbf J_0$ is the ground state of the Mexican-hat potential $U({\boldsymbol J})$, and we study the longitudinal fluctuations, $\delta J_\parallel$, and the perpendicular fluctuations, $\delta J_\perp$.  }
	\label{fig:schematic}
\end{figure*}

We consider two particle densities $\phi_a(\mbf x, t)$ (where $a=\{1,2\}$ labels the species), with conserved dynamics 
\beq 
\label{phi1}
\begin{cases}
\partial_t \phi_a+\mbf\nabla \cdot {\mbf j}_a=0, \\ 
{\mbf j}_a =- {\mbf \nabla} \left[\mu_a- \alpha \varepsilon_{a b}  \phi_b  -K \nabla^2 \phi_a\right]  - \sqrt{2 \mathcal D} \,{ \mbf \xi_a}. 
\end{cases}
\eeq 
%
The parameter $\alpha$ characterizes the non-reciprocal interaction, and thus also the activity. We are using a convention in which $\alpha >0$ when species 1 chases after species 2 (see Fig. \ref{fig:schematic}{\em A}). $\varepsilon_{a b}$ is the fully anti-symmetric Levi-Civita matrix and we use Einstein summation convention. For simplicity, we assume the same damping coefficient for both species, and absorb it in the unit of time. We also assume the same stiffness $K$, hence excluding the possibility of a small-scale Turing instability \cite{frohoff2021suppression, frohoff2021localized}. 
$\mu_a =\del f/\del \phi_a$ is the chemical potential expressed as the derivative of the free-energy density $f$. We use $f = - \frac{1}{2}( \phi_a \phi_a) + \frac 14 ( \phi_a \phi_a)^2$, which is invariant under orthogonal transformations in the $\phi_1, \phi_2$ plane and promotes equilibrium phase separation into $\phi_a\phi_a= 1$, corresponding to $f_{\rm min}=-\frac{1}{4}$ (see Fig. \ref{fig:schematic}{\em B}) \cite{saha2022effervescent}. This choice allows exact analytical calculations, whose predictions are parameter-free \cite{saha2022effervescent, rana2023defect}. 
The noise sources are not correlated between species and are characterized by zero mean and unit variance $\langle  \xi_{a,i}(\mbf x,t) \xi_{b,j}(\mbf x',t')\rangle =  \delta_{a b} \delta_{i j} \delta^d(\mbf x- \mbf x')\delta(t-t')$. The amplitude $\mathcal D$ is chosen to be the same for both densities. 

\subsection*{Travelling bands} The above form of NRCH model admits solutions in the form of travelling density waves (Fig. \ref{fig:schematic}{\em A}), which can be represented via a complex field defined as $\phi\equiv \phi_1+i \phi_2=|\phi| \, e^{i \left({\mbf q}_0 \cdot \mbf x-\omega_0 t\right)}$, with $|\phi|=\sqrt{1-K q_0^2}$ and $\omega_0=\alpha q_0^2$ \rev{(see {\em Materials and Methods} for a discussion on the possible values of $q_0$).} The solution is stable, and is associated with a cyclic orbit in the configuration space at any point $\mbf x$ that does not reside at the bottom of the free energy landscape, corresponding to $f=-\frac{1}{4} \left(1-K^2 q_0^4\right)$ and $\frac{K}{2} (\mbf \nabla \phi)^2+f=-\frac{1}{4} \left(1-K q_0^2\right)^2$ (see Fig. \ref{fig:schematic}{\em B}). In these solutions, parity and time-reversal symmetries, together with time- and space-translation symmetries are spontaneously broken. Moreover, an {\em emergent polar order} is observed in the system that is composed of two scalar fields. 

\subsection*{Polar order parameter} To understand the nature of the emergent polar order, we start by defining a local polar order parameter $\mbf J = \varepsilon_{a b} \phi_a \mbf \nabla \phi_b$, which is nonzero when the density waves have a phase difference, leading to parity symmetry breaking.  \rev{This quantity was introduced in \cite{saha2020scalar} where using numerical simulations} it was shown that $\mbf J$ transitions to taking non-vanishing values when $\alpha$ is tuned to exceed the coefficient of linear reciprocal interaction, which is set to zero here for simplicity. 

We derive the effective governing equation for the emergent polar order parameter $\mbf J(\mbf x,t)$, using the dynamical equations for the two concentrations Eq. \ref{phi1}. To the leading order \rev{(i.e. until second order in gradients)}, we obtain
\beq 
&& \partial_t \boldsymbol J  + \lambda_1 \boldsymbol J (\boldsymbol \nabla \cdot \boldsymbol J) + \lambda_2 (\boldsymbol J \cdot \boldsymbol \nabla) {\boldsymbol J}  = \Gamma \boldsymbol \nabla (\boldsymbol \nabla \cdot \boldsymbol J)   \nonumber \\ 
&& + D_{lJ} \mbf J( \mbf J \cdot \mbf \nabla)(\mbf \nabla \cdot \mbf J) + D_{mJ} \boldsymbol J \nabla^2 |\mbf J|^2 +  r_1 \boldsymbol J \nonumber \\ 
&& + \left( r- u|\boldsymbol J|^2\right)|\boldsymbol J|^2 \boldsymbol J - \boldsymbol\nabla P + \lambda_3 \boldsymbol J (\boldsymbol J \cdot \boldsymbol \nabla) \rho + {\mbf \xi}_J, \label{eomJ} 
\eeq 
where the different terms are organized such that the similarities to the Toner-Tu equations \cite{toner1995long} are highlighted. The quantities that are introduced in Eq.  \ref{eomJ} are functions of the amplitude $\rho(\mbf x,t) = \sqrt{\phi_a \phi_a}$, which for the traveling band solution will represent a uniform field with $\rho(\mbf x,t)=\rho_0=\sqrt{1-K q_0^2}$. To the lowest order, we find $\lambda_1=2 \alpha/ \rho^2$, $\lambda_2 = 2 \alpha/\rho^2$, 
$\Gamma=1-\rho^2-2 m$, $D_{lJ}= 4 K/ \rho^4$, $D_{mJ}= 2 K/ \rho^4$, $r_1= 2 (2 m + 5 \rho^2-3)\nabla^2 \rho/\rho$, 
$r= 2 (1-\rho^2)/\rho^4$, $u= 2 K / \rho^8$, $P= 2 \,\mbf J \cdot \boldsymbol{\nabla} m$, $\lambda_3= 4 \alpha/\rho^3$. Here, we have defined the auxiliary field $m=\rho^4(r-u|\mbf J|^2)/2$, which identically vanishes in the broken symmetry state. The noise term ${\mbf \xi}_J$ will be discussed below (see {\em SI Appendix} for the extended version of Eq.  \ref{eomJ}).


We can make a number of observations from Eq. \ref{eomJ}, which describes the collective dynamics of the emergent polar order parameter field. The coefficients of the symmetry-breaking advective terms, namely $\lambda_1$ and $\lambda_2$, are proportional to $\alpha$. It is thus evident that $\alpha$ assumes the role of the velocity of self-propulsion in the Toner-Tu description and breaks Galilean invariance \cite{toner1995long}. The deterministic dynamics of $\mbf J$ is not conserved; this is due to a dissipative force that can be expressed as the derivative of a Mexican hat potential $U(\mbf J)=-\frac{r}{4} |\mbf J|^4+\frac{u}{6} |\mbf J|^6$ (see Fig. \ref{fig:schematic}{\em C}), namely, $- \partial U(\mbf J) / \partial \mbf J =( r - u |\mbf J|^2)\mbf J |\mbf J|^2  $. For a uniform background amplitude $\rho_0$, the dissipative force vanishes at the steady-state value of $J_0=|\mbf J_0| =  \sqrt{r_0/u_0}$ when $r_0 > 0$ (Fig. \ref{fig:schematic}{\em C}), with $r_0=r(\rho_0) = 2(1- \rho_0^2)/\rho_0^4$ and $u_0 = u(\rho_0)=2 K/\rho_0^8$, namely, $J_0=\rho_0^2\sqrt{(1-\rho_0^2)/K}$. (Throughout the paper, the subscript $0$ always indicates that the quantity is evaluated at the background solution $\rho_0$). If we take the uniform amplitude state to represent the stable traveling band solution, then $J_0=\rho_0^2 q_0$. In dimensions $d >1$, this potential is characterised by rotational symmetry, which is broken when the ground state $\mbf J_0=J_0 {\hat {\mbf e}}$ spontaneously chooses a specific direction along ${\hat {\mbf e}}$.

To complete the effective description of the dynamics, we need a governing equation for the amplitude, which plays the role of the density in the Toner-Tu analogy. We find
\beq
&&\partial_t \rho +  \lambda_4(\mbf \nabla \cdot \mbf J) =\beta m |\mbf J|^2+D_\rho\nabla^2 \rho  \nonumber \\
&& + \kappa \nabla^2 |\mbf J|^2 + w (\mbf J \cdot \nabla)(\nabla \cdot \mbf J) + \xi_\rho,\label{eomRho}
\eeq 
with the coefficients given as $\lambda_4= \alpha/\rho$, $\beta=1/\rho^3$, $D_\rho=5- 3\rho^2 +2 m  \rev{+ \mathcal D/\rho^2}$, $\kappa=K/ \rho^3$, $w=2 K/\rho^3$, and the noise term $\xi_\rho$ will be discussed below (see {\em SI Appendix} for the extended version of Eq. \ref{eomRho}). 
We observe that the amplitude equation exhibits significant differences with both the conserved \cite{toner1995long} and the non-conserved \cite{Toner2012PRL} versions of Toner-Tu equations.

The noise terms $\boldsymbol \xi_J$ and $\xi_\rho$ can be derived straightforwardly from the stochastic sources of Eq.  \ref{phi1}, leading to
\begin{equation}
\label{noise}
\begin{gathered}
\boldsymbol \xi_J = \sqrt{2 \mathcal D} \varepsilon_{a b}( \phi_a \mbf \nabla - \mbf \nabla \phi_a)(\mbf \nabla \cdot \boldsymbol \xi_b),  \\
 \xi_\rho =  \sqrt{2 \mathcal D} (\phi_a/\rho) (\mbf \nabla \cdot \boldsymbol \xi_a),
\end{gathered}
\end{equation}
which are multiplicative noises and have the potential to be conserved (see {\em Materials and Methods}), in apparent contradistinction to the non-conserved noise in Toner-Tu equations \cite{toner1995long}. When we evaluate the noise terms using the traveling band solution, we obtain effective Gaussian noises with zero mean and the following correlators in Fourier space (see {\em Materials and Methods})
\beq
\label{noise-2}
\begin{gathered}
\langle  \xi_{J_{\perp}i}(\mbf q,\omega) \xi_{J_{\perp}j}(\mbf q',\omega')\rangle = 2 \mathcal D \rho_0^2 q_0^2 q_{\perp i} q_{\perp j} \,\delta_{\mbf q+ \mbf q'} \delta_{\omega+\omega'}, \\
\langle  \xi_{J_{\parallel}}(\mbf q,\omega) \xi_{J_{\parallel}}(\mbf q',\omega')\rangle = 8 \mathcal D \rho_0^2 q_0^4 \,\delta_{\mbf q+ \mbf q'} \delta_{\omega+\omega'}, \\
\langle  \xi_{\rho}(\mbf q,\omega) \xi_{\rho}(\mbf q',\omega')\rangle = 2 \mathcal D q_0^2 \,\delta_{\mbf q+ \mbf q'} \delta_{\omega+\omega'} \\
\langle  \xi_{\rho}(\mbf q,\omega) \xi_{J_\parallel}(\mbf q',\omega')\rangle = 4 \mathcal D \rho_0 q_0^3 \,\delta_{\mbf q+ \mbf q'} \delta_{\omega+\omega'},
\end{gathered}
\eeq
to the lowest order, where ${\mbf q}_{\perp}={\mbf q}-( {\mbf q} \cdot {\hat {\mbf e}}) {\hat {\mbf e}} $ and the short-hands $\delta_{\mbf q+ \mbf q'} \equiv (2 \pi)^{d} \delta^d(\mbf q+ \mbf q')$ and $\delta_{\omega+\omega'} \equiv (2 \pi)\delta(\omega+\omega')$ have been used. We thus observe that while the noise terms for the longitudinal emergent polar order parameter and the amplitude are non-conserved, the noise terms for the transverse emergent polar order parameter are conserved. 

\subsection*{Linear theory} The ordered state that is predicted by Eq.  \ref{eomJ} identifies a phase separated state with spatial modulation, as well as spontaneous breaking of time-reversal and rotational symmetry. To test the robustness of $\mbf J_0$ in the presence of noise, we linearly expand Eqs. \ref{eomJ} and \ref{eomRho} around the steady-state. We substitute $\boldsymbol J = \boldsymbol J_0 + \delta \boldsymbol J = (J_0 + \delta J_\parallel) {\hat {\mbf e}} + \delta \boldsymbol{J}_\perp$ distinguishing longitudinal and perpendicular fluctuations, and perturb the amplitude as $\rho = \rho_0 + \delta \rho$. We derive the fluctuating linear dynamics up to second order in gradients,
\begin{widetext}
\beq
\label{rho_fluctuations}
&&  \partial_t \delta \rho = -2 \beta_0 q_0^2 \rho_0^5  \gamma^2 \delta \rho - 2 \beta_0 \Gamma_{0} J_0 \delta J_\parallel - \lambda_{4,0}(\mbf \nabla \cdot \delta \boldsymbol J) + D_{\rho,0} \nabla^2 \delta \rho + w_0 J_0[\nabla^2 \delta J_\parallel+ \partial_\parallel (\mbf \nabla \cdot \delta \mbf J)] + {\xi}_\rho, \\
\label{parallel_fluctuations}
&&\partial_t \delta J_\parallel = - 4 q_0^2 \Gamma_0 \delta J_\parallel - 4  q_0^3 \rho_0^3 \gamma^2 \delta \rho - 2 J_0 \lambda_{1,0} (\partial_\parallel \delta J_\parallel) - J_0 \lambda_{1,0} (\mbf \nabla_\perp \cdot \delta \boldsymbol{J}_\perp) + J_0^2 \lambda_{3,0} \partial_\parallel \delta \rho + 2 q_0 \rho_0 D_{\rho,0} \nabla_\perp^2 \delta \rho  \\ \nonumber
&& \qquad + 2 q_0 \rho_0 (D_{\rho,0}+\rho_0^2 \gamma^2) \partial_\parallel^2 \delta \rho + 13 \Gamma_0 (\partial_\parallel^2 \delta J_\parallel) + 4 \Gamma_0 \nabla_\perp^2 \delta J_\parallel + 5 \Gamma_0  \partial_\parallel(\mbf \nabla_\perp \cdot \delta \boldsymbol{J}_\perp) + {\xi}_{J_\parallel}, \\
\label{perpendicular_fluctuations}
&& \partial_t \delta \boldsymbol J_\perp = - \lambda_{2,0} J_0 \partial_\parallel \delta \boldsymbol J_\perp  + 4  q_0 \rho_0^3 \gamma^2 \mbf \nabla_\perp \partial_\parallel \delta \rho + \Gamma_0 \mbf \nabla_\perp(\mbf \nabla_\perp \cdot \delta \boldsymbol{J}_\perp) + 5 \Gamma_0 \mbf \nabla_\perp ( \partial_\parallel \delta J_\parallel)+ {\boldsymbol \xi}_{J_\perp},
\eeq 
\end{widetext}
where we have used $\partial_\parallel \equiv {\hat {\mbf e}} \cdot \mbf \nabla$, $\mbf \nabla_\perp \equiv \mbf \nabla -{\hat {\mbf e}} ({\hat {\mbf e}} \cdot \mbf \nabla)$, and $\gamma^2= (3 \rho_0^2 -2)/\rho_0^2$. 

\subsection*{The slow modes} From Eq.  \ref{perpendicular_fluctuations},  the fluctuations that are perpendicular to the broken symmetry direction, $\delta \mbf J_\perp$, can be identified as slow modes of the model, as they represent the Goldstone modes associated with the continuous rotational symmetry breaking (Fig \ref{fig:schematic}). While $\delta \rho$ and $\delta J_\parallel$ appear to be fast variables, the existence of a constraint in the form of a curl-free condition $\partial_i (J_j/\rho^2) - \partial_j(J_i/\rho^2)=0$ suggests that an additional slow mode that combines the two fields also exists in the dynamics. Examining the eigen-mode structure of Eqs. \ref{rho_fluctuations} and \ref{parallel_fluctuations}, we identify the new slow variable as
$\delta s = \delta J_\parallel-2 (J_0/\rho_0) \delta \rho$. 
Solving for different linear slow modes, we find
\beq
&&\delta {\mbf J}_{\perp}(\mbf q,\omega)={\cal G}(\mbf q,\omega) \,{\mbf \xi}_{J_{\perp}}(\mbf q,\omega), \label{lineardyn} \\ 
&&\delta s(\mbf q,\omega)={\cal G}(\mbf q,\omega) \,\xi_{s}(\mbf q,\omega) , \label{lineardyn-2}
\eeq
in terms of the Green function of the linear dynamics ${\cal G}(\mbf q,\omega)=\left[i(\omega-v_g q_{\parallel})+\Gamma_0 \left(q_\perp^2 + \gamma^2 q_\parallel^2\right)\right]^{-1}$. 
The coefficient $v_g = 2 \alpha q_0$ advects the fluctuations along the ordering direction and represents the propagating sound speed. The diffusion is anisotropic with coefficients $\Gamma_0=K q_0^2$ along the transverse directions and $\Gamma_0 \gamma^2=K q_0^2 (3\rho_0^2-2)/\rho_0^2$ along the longitudinal direction. Note that the stability of the dynamics requires $\rho_0^2 > 2/3$ (in order to have $\gamma^2 >0$), which is connected to the Eckhaus instability \cite{aranson2002world}. The noise for the new slow mode in Eq.  \ref{lineardyn-2} has zero mean and the following correlator (see {\em Materials and Methods}) 
\beq
\langle  \xi_{s}(\mbf q,\omega) \xi_{s}(\mbf q',\omega')\rangle = 2 \mathcal D \rho_0^2 q_0^2 q_{\parallel}^2 \,\delta_{\mbf q+ \mbf q'} \delta_{\omega+\omega'}.
\eeq
We note that the dynamics of the slow modes corresponds to a rather uncommon class where the deterministic evolution of the dynamics is dissipative while the noise contribution that drives the stochastic fluctuations is conserved \cite{hohenberg1977theory}. 

\subsection*{Long-Range Oder in $d>1$} We can now calculate the magnitude of the transverse fluctuations of the emergent polar order parameter. Combining Eqs.  \ref{lineardyn} and \ref{noise-2}, we have $\langle |\delta \mbf J_\perp (\boldsymbol x,t)|^2 \rangle=\int_\omega \int^\Lambda_{\mbf{q}} 2 \mathcal D \rho_0^2 q_0^2 q_{\perp}^2 |{\cal G}(\mbf q,\omega)|^2$ with $\Lambda$ being the finite ultraviolet cutoff. The integrals can be calculated to give
\beq 
\label{LRO}
\langle |\delta \mbf J_\perp (\boldsymbol x,t)|^2 \rangle =(d-1) \frac{\mathcal D \rho_0^2}{d K}  \frac{\Lambda^d}{(2\pi)^d}, 
\eeq 
which stays finite below the onset of Eckhaus instability for any $d >1$. Therefore, in our theory transverse fluctuations are suppressed and the system exhibits true long-range order in $\mbf J$ for any $d>1$ even at the linear level. 

\subsection*{Non-linear theory} To verify to which extent the predictions of linear theory hold, we need to understand the role of non-linearities. The most relevant contributions are given by terms containing one gradient and two fields, coupling the dynamics of transverse and longitudinal slow modes. Below, we will show that these are the solely relevant nonlinear term as predicted by the Renormalization Group framework. After expanding Eqs.  \ref{eomJ} and \ref{eomRho} at this order in fluctuations, changing to the comoving frame of reference (with velocity $v_g \, {\hat {\mbf e}}$), and rescaling the longitudinal coordinate by $\gamma$, we can write down the dynamics for a vector field $\mbf p =\gamma \,\delta s \, {\hat {\mbf e}}+\delta \mbf J_\perp$, as follows
\beq 
\label{Peq}
\partial_t \mbf p + \lambda_0 \mbf p \cdot \mbf \nabla \mbf p = \Gamma_0 \nabla^2 \mbf p - \mbf \nabla \eta(\mbf x,t),
\eeq 
where the nonlinear coupling constant can be traced back to the non-reciprocity as $\lambda_0 =2 \alpha/\rho_0^2$, and we have used the curl-free condition reported above as inherited by $\mbf p$, namely $\partial_i p_j - \partial_j p_i=0$. The variance of the noise reads $\langle \eta(\mbf q,\omega) \eta(\mbf q',\omega')\rangle = 2 D_0 \,\delta_{\mbf q+ \mbf q'} \delta_{\omega+\omega'}$, where the amplitude is given as $D_0=\mathcal D q_0^2 \rho_0^2 \gamma$. Remarkably, we thus obtain the well-known noisy Burgers equation \cite{forster1977large, medina1989burgers}. 

We finally express the polar vector field as the gradient of a scalar function $\mbf p =  -\mbf \nabla h$, as the curl-free constraint implies \cite{medina1989burgers}. The dynamics Eq. \ref{Peq} thus becomes
\beq 
\partial_t h = \Gamma_0 \nabla^2 h + \frac{\lambda_0}{2} (\nabla h)^2 + \eta (\mbf x,t),
\eeq 
namely, the celebrated Kardar-Parisi-Zhang (KPZ) equation in any dimension $d$ \cite{kardar1986dynamic}. This result states that the non-linear dynamics of the fluctuating modes around the ordered traveling state of our system can be mapped to the equation for growing interfaces described by the height function $h(\mbf x,t)$, thus belonging to its universality class. Non-reciprocity, here described by the parameter $\alpha$, is the key ingredient to connect the NRCH model for particle densities to the KPZ equation. We note that the KPZ field represents the fluctuations of the constant phase manifolds in the underlying complex field theory involving the order parameter $\phi=\rho \, e^{i \theta}$, namely, $\theta(\mbf x,t)=\gamma {\mbf q}_0 \cdot \mbf x-(\omega_0-v_g q_0) t-h(\mbf x,t)$ in terms of the new coordinates. Therefore, the flatness or roughness of the KPZ field can be interpreted as a reflection on the shape of the bands in the NRCH model, which effectively represents an active traveling smectic phase \cite{saha2020scalar}.

\begin{figure*}[t]
    \includegraphics[width=\textwidth]{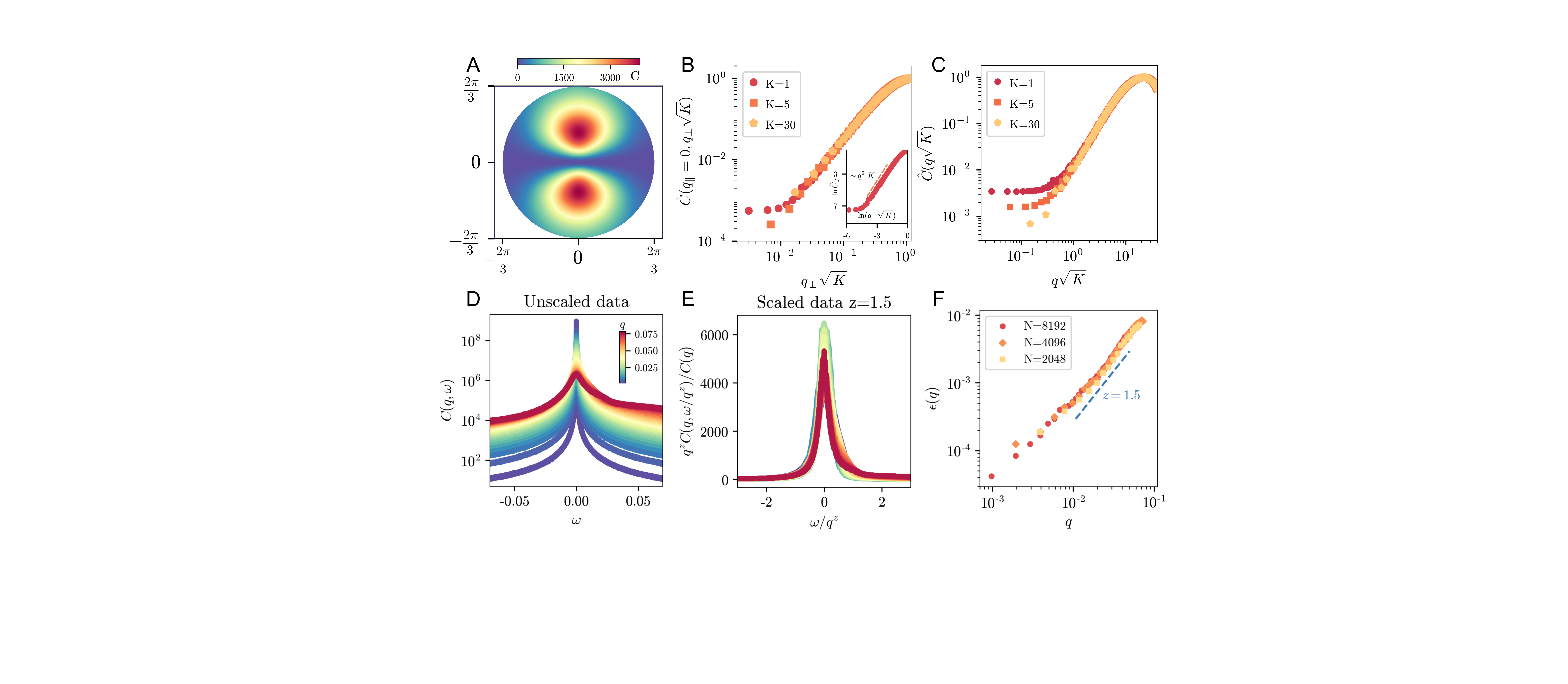}
    \caption{\rev{Numerical simulations of Eq. \ref{phi1}. 
    {\em (A)}
    Static correlation function of $\delta J_\perp$ in $d=2$ in full momentum space corresponding to the first Brillouin zone. 
    {\em (B)}
    A cut of the correlation function presented in panel {\em A} as normalized by the maximum value $\hat{C}=C/C_\text{max}$ at $q_\parallel=0$ for different values of $K$. Inset: scaling of $\sim q_\perp^2$ due to the conserved noise. 
    {\em (C)}
    Static correlation function of $\delta s$ in $d=1$ at different values of $K$. Both in $d=2$ and $d=1$ we find a plateau at small $q$ describing the property of long-range order in the thermodynamic limit. Parameters of simulations: $\alpha=1, {\mathcal D}=0.1, K=1, q_0=2 \pi/L$, in $d=2$ $N=2048, L=2048$ and in $d=1$, $N =4096, L=4096$. 
    {\em (D)}
    Shifted dynamical correlation function of $\delta s$ in frequency space at different wave-numbers for $d=1$. The peaks at different speed $v_g$ are shifted to the reference frame of the traveling band. 
    {\em (E)}
    Characteristic collapse of curves of panel {\em D} when the dynamical scaling hypothesis is applied with the exponent $z=3/2$, corresponding to $d=1$.
    {\em (F)}
    Comparison between the analytical $z$ and the behavior of width at half-height of the correlation functions presented in panel {\em D}, as functions of wave-number, for $d=1$. Parameters of panels {\em D} and {\em E}: $N=8192, L=2048 \pi, \alpha=5, q_0=0.15625, K=1, {\mathcal D}=0.05, dt=0.01, t_\text{max}=\text{1-3} \times 10^5$. For smaller sizes, we have $N=4096,L=1024 \pi$, and, $N=2048,L=512 \pi$. These results are consistent with the KPZ physics in $d=1$. }}
    \label{fig:correlations}
\end{figure*}

\subsection*{Renormalization Group predictions} We can now use the \rev{known} results on the critical dynamics of the KPZ equation to characterize the scaling behavior of our emergent polar order parameter. For a scaling factor $b=e^l \gtrsim1$, we will seek to find scaling transformations in the form of as $ t \to b^z t,\ \mbf{x} \to b  \mbf x, \  \mbf p \to b^{\zeta-1}  \mbf p$ such that the dynamics Eq. \ref{Peq} is scale invariant \cite{medina1989burgers}. Here, $z$ is the dynamical critical exponent, and $\zeta$ is the roughness exponent of the underlying KPZ field $h$. Applying these transformations to Eq.  \ref{Peq}, we obtain the following scaling relations for the coupling constants of the field theory: $\Gamma_b = \Gamma_0 b^{z-2}$, $D_b= D_0 b^{z -d - 2 \zeta}$, and $\lambda_b = \lambda_0 b^{\zeta+z-2}$. The Gaussian fixed point corresponds to $\lambda^*=0$ at which the long-wavelength dynamics is ruled by the linear theory with exponents $z=2$ and $\zeta=1-d/2$. These exponents determine the critical dimension $d_c=2$ at which the non-linearity $\lambda_0$ is only marginally relevant. With these scaling dimensions, it is easy to verify that all higher order non-linearities are irrelevant. 

Perturbative Renormalization Group calculations performed for the effective coupling constant of theory $g=K_d\,{\lambda_0^2 D_0}/{\left(4 \Gamma_0^3\right)}$ (in which $K_d \equiv {S_d}/{(2 \pi)^d })$ and $S_d=2 \pi^{d/2}/\Gamma(d/2)$ is the area of the unit-sphere in $d$ dimensions) yields
\beq
\label{KPZ-flow}
\frac{{\rm d} g}{{\rm d} l}=(2-d) g+\frac{2}{d}\left(2 d-3\right) g^2,\label{RG-flow}
\eeq
at one-loop order \cite{medina1989burgers}. For $d=1$, Eq.  \ref{RG-flow} suggests that the dynamics is governed by a stable fixed point at $g^*=1/2$ (with the fixed point at $g^*=0$ being unstable), which corresponds to exponents $\zeta=1/2$ and $z=3/2$ that turn out to be exact. At $d=2$, \rev{the RG scenario is more delicate and deserves more attention \cite{Frey1994}}: the fixed point at $g^*=0$ is marginally unstable, which hints at the existence of a correlation length given as
\beq
\ell_{\times}=\frac{2 \pi}{q_0} \exp\left(\frac{2 \pi K^3 q_0^4}{\gamma \alpha^2 \mathcal{D}}\right).\label{correlation-length-2D}
\eeq
For length scales smaller than $\ell_{\times}$, the dynamics is governed by the $g^*=0$ fixed point that corresponds to $\zeta=0$ and $z=2$, whereas for length scales larger than $\ell_{\times}$ a strong coupling fixed point controls the dynamics \cite{canet2011nonperturbative}. For $d=3$, an unstable fixed point $g^*=1/2$ separates two phases: a flat phase characterized by the stable fixed point $g^*=0$ corresponding to $\alpha < \alpha_c$, and a rough phase controlled by a strong coupling fixed point corresponding to $\alpha > \alpha_c$. The onset of the roughening transition occurs at 
\beq
\alpha_c=\left(\frac{\pi^2 K^3 q_0^4}{\gamma \mathcal{D}}\right)^{1/2}.\label{alpha-c}
\eeq
Calculations up to two-loop order (that introduce higher order terms in Eq.  \ref{KPZ-flow}) do not significantly change the above conclusions  \cite{Frey1994,wiese1997critical}. Note that NRCH provides a unique opportunity for an experimental realization of the 3D KPZ universality class, and the corresponding roughening transition, without the need to have access to 4D position space.

It is important to examine how the question of true long-range emergent polar order is influenced by the presence of nonlinearity in the dynamics, and how it is affected by whether the constant-$\theta$ bands are statistically flat or rough. The fluctuations in the polarization can be calculated as
\beq
\langle {\mbf p} (\mbf x,t)^2\rangle =\langle {\mbf \nabla} h(\mbf x,t)^2\rangle \sim \int_{1/L}^{\Lambda} q^{1-2\zeta}d q \sim \Lambda^{2(1-\zeta)},\label{fluc-p-zeta}
\eeq
which is finite as long as $\zeta < 1$ holds; note that Eq.  \ref{fluc-p-zeta} would yield $\sim L^{2(\zeta-1)}$ when $\zeta >1$, which would diverge with the system size $L$. Inserting the Gaussian fixed point value of $\zeta=1-d/2$ in Eq.  \ref{fluc-p-zeta} yields $\Lambda^d$, which is the result reported in Eq.  \ref{LRO}. It is, indeed, known that $\zeta<1$ generally holds for KPZ equation in any dimension, with the most recent conjecture of $\zeta=7/(4d+10)$ (and, correspondingly, $z=\frac{8d+13}{4d+10}$) representing well the numerically obtained results so far \cite{KPZ-all-d}. Therefore, the true long-range order in the emergent polar order parameter persists even in the presence of the nonlinear term and when the underlying KPZ dynamics is governed by the perturbatively inaccessible strong coupling fixed point, e.g. in $d=2$ and beyond the roughening transition (corresponding to $\alpha > \alpha_c$) in $d=3$. 

\rev{We tested our results with numerical simulations of Eq. \ref{phi1} in the regime of weak noise. We have run simulations both in 1D and 2D. For the 2D case, we evaluate the equal-time correlation of the fluctuations in the direction perpendicular to the polar order $C(\mbf q)=\langle \delta J_\perp(\mbf q) \delta J_\perp(-\mbf q) \rangle$ (Fig. \ref{fig:correlations}{\em A}). We observe a saturation in the limit of small wave-numbers to a finite value (determined by simulation parameters). This is evident also in the cross section at $q_\parallel=0$ of Fig. \ref{fig:correlations}{\em B}, reflecting the analytical form and confirming our prediction on true long-range order in $d=2$. Moreover, at larger $\mbf q$, we see a quadratic growth due to the conserved noise, which scales as expected for different values of $K$. These properties are persistent also for the 1D case in the static correlation function of the longitudinal slow mode $\delta s$ (Fig. \ref{fig:correlations}{\em C}). We also investigated the dynamics of the system in 1D and tested the relevance of non-linear terms in the regime of large non-reciprocity. We have computed the dynamical correlation functions $C(q,\omega) = \langle \delta s(q, \omega) \delta s(-q,-\omega) \rangle$ at different wave-numbers and presented them in Fig. \ref{fig:correlations}{\em D} with a shift to the reference frame of the traveling waves $\omega \to \omega + v_g q$. We then test the dynamical scaling hypothesis using $z=3/2$, which is the dynamical exponent of the stable perturbative fixed point of the KPZ equation  \cite{medina1989burgers}. We observe a good collapse on the same shape-function for more than one decade of wave-numbers (Fig \ref{fig:correlations}{\em E}). Finally, we evaluate the width at half height $\epsilon(q)$ of the correlations in Fig. \ref{fig:correlations}{\em D}, and we compare its behavior with the expected analytical result (Fig. \ref{fig:correlations}{\em F}). This probe shows a very good agreement with the KPZ universality class. We note that the smallest wave-numbers show deviations from the epxected scaling due to the finite limit of the simulation time. We would like to highlight that this study is performed by evaluating fluctuations around a traveling wave state, which is a novel analysis from a methodological point of view (see {\em Materials and Methods}). }


\section*{Discussion and conclusions} We present a new effective theory for a mixture of two species with non-reciprocal interaction as described by conserved scalar fields, in terms of an emergent polar order parameter field that breaks time-reversal symmetry. Our framework shows striking similarities with the Toner-Tu theory of dry polar active matter, most notably an ordering potential and nonlinearities describing activity-driven advection. The effective theoretical framework for the emergent polar order parameter field predicts rotational symmetry breaking and the existence of Goldstone modes that emerge as a result of broken rotational symmetry. The theory, however, features marked differences with the Toner-Tu theory. The amplitude equation is not equivalent either to the conserved \cite{toner1995long} or the Malthusian \cite{Toner2012PRL} versions of Toner-Tu equations. Moreover, the noise that drives the soft modes is conserved, which leads to a violation of the Mermin-Wagner theorem in $d=2$ \cite{binney1992theory}, already at the linear level, as the low-cost and thus easily excitable elastic deformations of the Goldstone modes are suppressed by the vanishing strength of the spontaneous fluctuations at the largest length scales. We note that most existing theories of polar active matter do not show long-range order in $d=2$ at the linear level, and non-linear active terms are necessary to tame fluctuations around the ordered state \cite{toner1998flocks}, with the exception of theories that incorporate a momentum conserving fluid near a boundary  \cite{uchida2010synchronization,uchida2010synchronization2,sarkar2021swarming}. 

The predictions of the linear theory are validated by our analysis including nonlinear terms. We show that the fluctuating modes of our theory follow a noisy Burgers equation for a single curl-free vectorial field, which can be mapped to a KPZ dynamics in every $d$. We observe that the relevant nonlinearity is produced by non-reciprocity and cannot generate any non-conserved noise term under renormalization. Building on the effective KPZ description of the fully nonlinear theory, we prove that the system exhibits true long-range polar order in any dimension, which is the central result of our work. \rev{Our analytical results are confirmed by our numerical simulations of the original NRCH model.}

We would like to close by highlighting an important feature of the emergent polar order parameter, which can be written as $\mbf J=\frac{1}{2i} \left(\phi^* \mbf \nabla \phi-\phi \mbf \nabla \phi^*\right)$ in analogy to quantum mechanics: it has been constructed to measure the coherence between the two species in the NRCH model. In light of this definition, one can argue that investigating the dynamics of $\mbf J$ follows the same spirit as studying the effective dynamics of composite particles in quantum condensed matter systems \cite{BCS1957,LeeRMP2006}. Moreover, since coherence is the interesting physical observable, $\phi$ plays a role that is more analogous to a wave function than a density, whereas $\rho^2$ plays the role of density or probability, again, highlighting the significance of the composite particles that chase each other taking on the role of the fundamental unit of the effective theory, leading to the emergence of an unanticipated polar symmetry.

\section*{Matherials and Methods}
\subsection*{Numerical Simulations}
\label{sec:Numerical}

Simulations shown in Fig. \ref{fig:schematic}{\em A} \rev{and Fig. \ref{fig:correlations}} have been performed using a pseudo-spectral method with periodic boundary conditions. The algorithm combines the evaluation of linear terms in Fourier space and non-linear terms in real space in order to obtain a stable solution of the non-linear partial differential equations (PDE)s; more details can be found in Ref. \cite{saha2020scalar}. 
 We use a backward Euler-Maruyama method to perform the time integration \cite{talay1994numerical}. 
The noise fields are generated at each point of the lattice and each time-step from a Gaussian distribution with zero mean and unit width. 

\rev{ {\it Static correlation functions}. We initialize the system with a periodic pattern of minimum wave-number $ q_{\rm min} = 2 \pi /L$ along the $x$ direction. After a transient period of thermalization, we compute the fluctuations $\delta J_\perp (\mbf x,t)$ and $\delta s(\mbf x,t)$, and we measure the correlations in Fourier space by averaging over time. 

{\it Dynamical correlation functions.} We perform this study in $d=1$. To maximize a good resolution in $\omega$ space for the dynamics, we initialize the system with a larger value of $q_0$, which we chose as $q_0=0.15625$. This is relevant for obtaining the results shown in Fig. \ref{fig:correlations} for two practical reasons. Firstly, the inverse of $q_0$ sets a minimum length scale under which we cannot evaluate the scaling  hypothesis; therefore, we have to restrict to $q<q_0$ to capture the collective effects. Secondly, as the analytical calculations show, the value of $q_0$ crucially determines the width of the bare dynamical correlation functions through $\Gamma_0$ as well as the value of the coupling constant $g$. 

We probe fluctuations in frequency and wave-vector space $\delta s(q,\omega)$. To ensure correct results in frequency space, we apply a windowing procedure with a Hanning function to the non-periodic time signals. We then compute $C(q,\omega) = \langle \delta s(q, \omega) \delta s(-q,-\omega) \rangle$ and average over $800$ independent noise realizations. The results are curves which peak at $\omega = v_g q$, which represents the sound mode of our system. Before testing the dynamical scaling hypothesis, we shift the curves to this value, so that they are symmetric around zero and we can focus on the scaling of the characteristic frequency. This procedure corresponds to the passage to the co-moving frame used in the analytical calculations. 
After the shifting we can test the dynamical scaling hypothesis \cite{halperin1969scaling}:
\beq 
\frac{C(q,\omega)}{C(q)}  \sim \frac{1}{q^z} {\cal F}( \omega/q^z),
\eeq 
where ${\cal F}$ is a universal shape-function. 
Note that to have a good resolution in $\omega$ space, we need to run long simulations at large $t_\text{max}$, which becomes critical for resolving the narrow divergence for small wave-numbers. 
}

 \subsection*{Noise for the slow mode dynamics}
\label{sec:noise_in_methods}
Equation \ref{noise} represents the noise corresponding to the polar order parameter and the amplitude, as derived from the conserved additive noise of Eq.  \ref{phi1}. \rev{Since the derivation of our effective field theory involved analytical manipulations of a field theory that is described in terms of one set of fields and the corresponding coordinate transformation to another set of fields, which involves frequent usage of the chain rule, we chose to perform our calculations using the Stratonovich convention, to be able to implement the standard chain rule for derivatives without the need for an additional term arising form the Ito calculus \cite{gardiner1985handbook}. However, we also checked that the results are not affected by this choice as expected.}

In order to determine the relevant noise contributions to the linear dynamics of $\delta \rho$ and $\delta \mbf J$, we expand the conserved multiplicative noise of Eq. \ref{noise} around the traveling wave state $ \phi_a = \bar{\phi}_a( \mbf x ,t) + \delta \phi_a( \boldsymbol x,t)$, with  $\bar{\phi}_1 = \rho_0 \cos \theta_0(\mbf x,t)$ and 
$\bar{\phi}_2 = \rho_0 \sin \theta_0(\mbf x,t)$.  For fluctuations of the polar order parameter we obtain
\beq 
&& \mbf \xi_J = \sqrt{2 \mathcal D} [ \varepsilon_{ab} \mbf \nabla \bar{\phi}_b \mbf \nabla \cdot \boldsymbol \xi_a + \varepsilon_{ab} \bar{\phi}_a \mbf \nabla (\mbf \nabla \cdot \boldsymbol \xi_b)],\nonumber
\eeq 
which we can subsequently project onto the perpendicular and longitudinal directions, as follows
\begin{eqnarray}
&&\xi_{J_\parallel} = \sqrt{2 \mathcal D }  [ \varepsilon_{ab} (\partial_\parallel \bar{\phi}_b) \mbf \nabla \cdot \boldsymbol \xi_j + \varepsilon_{ab} \bar{\phi}_a \partial_\parallel (\mbf \nabla \cdot \boldsymbol \xi_j)] , \nonumber \\
&&\mbf \xi_{J_\perp} =  \sqrt{2 \mathcal D} \varepsilon_{ab} \bar{\phi_a} \mbf \nabla_\perp (\mbf \nabla \cdot \boldsymbol \xi_b) ,\nonumber
\end{eqnarray}
while for the amplitude fluctuations we obtain
\beq 
\label{deltarhonoise}
\xi_\rho = \sqrt{2 \mathcal D}(\bar{\phi}_a/\rho_0)  \mbf \nabla \cdot \boldsymbol \xi_a.\nonumber
\eeq
In Fourier space, the main effect of the spatio-temporal oscillations is to translate wave-number and frequency by the selected $q_0$ and $\omega_0$. Mean values are null, and the variances for the longitudinal fast fields become,
\beq
&& \langle \xi_\rho(\mbf q,  \omega) \xi_\rho(\mbf q',  \omega') \rangle = 2 \mathcal D (q_0^2+q^2) \delta_{\mbf q+\mbf q'} \delta_{\omega+\omega'}, \nonumber \\
&& \langle \xi_{J\parallel}(\mbf q, \omega) \xi_{J\parallel}(\mbf q', \omega') \rangle = 8 \mathcal D \rho_0^2 q_0^4 \delta_{\mbf q+\mbf q'} \delta_{\omega+\omega'} + \nonumber \\ 
&& + 2 \mathcal D \rho_0^2 q_0^2 (4 q^2 + 9 q_\parallel^2 + q^2 q_\parallel^2/q_0^2 )\delta_{\mbf q+\mbf q'} \delta_{\omega+\omega'},\nonumber
\eeq
 We note that the leading contributions are non-conserved. However, they cancel in the definition of $\delta s$ producing conserved noise for the longitudinal and transverse slow modes:
\begin{align}
   &\langle \xi_s(\mbf q, \omega) \xi_s(\mbf q', \omega') \rangle =  2 \mathcal D \rho_0^2 q_\parallel^2 ( q_0^2 +q^2 ) \delta_{\mbf q+\mbf q'} \delta_{\omega+\omega'}, \nonumber \\
    & \langle \xi_{J_{\perp} i}( \mbf q, \omega) \xi_{J_{\perp} j}(\mbf q', \omega') \rangle =  2 \mathcal D \rho_0^2 q_{\perp i} q_{\perp j} \left( q_0^2 +q^2 \right)\delta_{\mbf q+\mbf q'} \delta_{\omega+\omega'}. \nonumber \\ \nonumber
\end{align}
At the leading order, these are conserved additive noise terms with amplitudes of order $\sim q^2$. At large momenta the conservation law produces an order $\sim q^4$ behaviour.

\rev{\subsection*{Wave-number selection} We assumed the existence of a finite wave-number $\mbf q_0$ throughout our calculations. This assumption is justified by our observation of a wave-number selection mechanism when non-reciprocity takes over and the ordered pattern is formed (see numerical simulations of \cite{saha2020scalar,rana2023defect}). 
To understand this observation, we can perform a calculation in which we rewrite Eq. \ref{phi1} in complex notation $\phi=\phi_1+i \phi_2$, and then solve for the dynamics for the structure factor $S_{\mbf q}(t) = \langle \phi_{\mbf q}(t) \phi^*_{\mbf q}(t) \rangle$. At the linear level, we obtain
\beq 
\partial_t S_{\mbf q}(t) = 2 q^2 (1 - K q^2)S_{\mbf q}(t) = \mathcal L(q) S_{\mbf q}(t).
\eeq 
Importantly, the function $\mathcal L(q)$ goes to zero for $q \to 0$ because of the particle number conservation law. The maximum of $\mathcal L(q)$, namely $q_\text{max}= 1/\sqrt{2 K}$, corresponds to the fastest growing mode, which is therefore finite. In the absence of the conservation law, the maximum would be at vanishing $q$, meaning that the system relaxes to the minimum wave-number $\sim 1/L$ \cite{montagne1996numerical}. However, the linearly predicted value $q_\text{max}$ is above the Eckhaus threshold \cite{aranson2002world} and hence not stable. We expect the non-linearities to lower the value of the spontaneously selected wave-number to a value that lies in the stable region, while the conservation law would ensure that it would stay finite in the infinite size limit.}

\rev{\subsection*{Justification for the number of slow modes}
The model of Eq. \ref{phi1} describes two conserved particle densities, which are therefore slow modes of the dynamics. When performing the calculation to rewrite the linearized dynamics in terms of the amplitude and the phase of the traveling band solution \eqref{rho_fluctuations},\eqref{parallel_fluctuations}, \eqref{perpendicular_fluctuations}, we have made the assumption to consider a constant phase shift $\Delta=\pi/2$ between the $\phi_1$ and $\phi_2$ oscillating profiles. 

To be certain to have included all the relevant slow modes of the theory, we can relax this constraint and analyze the dynamics of phase shift fluctuations. To simplify our argument, we neglect the amplitude fluctuations and consider a general solution in the form of two coexisting smectics with two independent displacement fields, namely
 \beq 
&& \phi_1(\mbf x,t) = \frac12 \left[\rho_0 e^{i \theta_0(\mbf x,t)+ i q_0 u_1(\mbf x,t)} + \text{c.c.}\right],\\ 
&&  \phi_2(\mbf x,t) = \frac12 \left[ \rho_0 e^{i \theta_0(\mbf x,t) + i q_0 u_2(\mbf x,t)} + \text{c.c.} \right],\nonumber
\eeq 
where $\theta_0= \mbf q_0 \cdot \mbf x - \omega_0 t$, and $\phi_0= \rho_0 e^{i \theta_0}$ is the background solution corresponding to $J_0$. In these terms, $\Delta(\mbf x,t) = q_0 (u_1- u_2)$ represents the relative phase fluctuations between the two smectics layers, while $q_0 u_1 = h( \mbf x,t)$. We thus write $\Delta(\mbf x,t) = \pi/2 + \delta \Delta(\mbf x,t)$ and derive the dynamics of the phase shift fluctuations by neglecting the coupling with $h$. At zeroth order in gradients, we obtain
\beq 
&& \partial_t \delta \Delta = - \left(\alpha  q_0^2  \tan \theta_0\right) \delta \Delta - \rho_0^2 q_0^2 \left( 1 - \cos 2 \theta_0\right) \delta \Delta,   
\eeq 
after dividing both sides of the equation by $\cos \theta_0$. Assuming that the fluctuations relax on temporal scales that are sufficiently longer than the period of the oscillations of the background phase $\theta_0$, we can further use a separation of time scale. Integrating over the fast oscillations yields 
\beq 
\partial_t \delta \Delta =   - \rho_0^2 q_0^2   \delta \Delta.
\eeq 
We thus conclude that the additional mode related to phase-shift fluctuations is a fast mode of the dynamics, which can therefore be neglected as it does not change the long-wavelength physics of the system presented in the main text. 
}

\subsection*{Data, Materials, and Software Availability} All study data are included in the article and/or supporting information.
\newline

We acknowledge discussions with Jaime Agudo-Canalejo, Benoît Mahault, Martin Johnsrud, Ahandeep Manna, Navdeep Rana, and Jacopo Romano. This work has received support from the Max Planck School Matter to Life and the MaxSyn-Bio Consortium, which are jointly funded by the Federal Ministry of Education and Research (BMBF) of Germany, and the Max Planck Society.

\bibliographystyle{apsrev4-2}  
\bibliography{biblio_NRCH2} 

\begin{thebibliography}{52}%
\makeatletter
\providecommand \@ifxundefined [1]{%
 \@ifx{#1\undefined}
}%
\providecommand \@ifnum [1]{%
 \ifnum #1\expandafter \@firstoftwo
 \else \expandafter \@secondoftwo
 \fi
}%
\providecommand \@ifx [1]{%
 \ifx #1\expandafter \@firstoftwo
 \else \expandafter \@secondoftwo
 \fi
}%
\providecommand \natexlab [1]{#1}%
\providecommand \enquote  [1]{``#1''}%
\providecommand \bibnamefont  [1]{#1}%
\providecommand \bibfnamefont [1]{#1}%
\providecommand \citenamefont [1]{#1}%
\providecommand \href@noop [0]{\@secondoftwo}%
\providecommand \href [0]{\begingroup \@sanitize@url \@href}%
\providecommand \@href[1]{\@@startlink{#1}\@@href}%
\providecommand \@@href[1]{\endgroup#1\@@endlink}%
\providecommand \@sanitize@url [0]{\catcode `\\12\catcode `\$12\catcode
  `\&12\catcode `\#12\catcode `\^12\catcode `\_12\catcode `\%12\relax}%
\providecommand \@@startlink[1]{}%
\providecommand \@@endlink[0]{}%
\providecommand \url  [0]{\begingroup\@sanitize@url \@url }%
\providecommand \@url [1]{\endgroup\@href {#1}{\urlprefix }}%
\providecommand \urlprefix  [0]{URL }%
\providecommand \Eprint [0]{\href }%
\providecommand \doibase [0]{https://doi.org/}%
\providecommand \selectlanguage [0]{\@gobble}%
\providecommand \bibinfo  [0]{\@secondoftwo}%
\providecommand \bibfield  [0]{\@secondoftwo}%
\providecommand \translation [1]{[#1]}%
\providecommand \BibitemOpen [0]{}%
\providecommand \bibitemStop [0]{}%
\providecommand \bibitemNoStop [0]{.\EOS\space}%
\providecommand \EOS [0]{\spacefactor3000\relax}%
\providecommand \BibitemShut  [1]{\csname bibitem#1\endcsname}%
\let\auto@bib@innerbib\@empty
\bibitem [{\citenamefont {Gompper}\ \emph {et~al.}(2020)\citenamefont
  {Gompper}, \citenamefont {Winkler}, \citenamefont {Speck}, \citenamefont
  {Solon}, \citenamefont {Nardini}, \citenamefont {Peruani}, \citenamefont
  {L{\"o}wen}, \citenamefont {Golestanian}, \citenamefont {Kaupp},
  \citenamefont {Alvarez}, \citenamefont {Ki{\o}rboe}, \citenamefont {Lauga},
  \citenamefont {Poon}, \citenamefont {DeSimone}, \citenamefont
  {Mui{\~{n}}os-Landin}, \citenamefont {Fischer}, \citenamefont {S{\"o}ker},
  \citenamefont {Cichos}, \citenamefont {Kapral}, \citenamefont {Gaspard},
  \citenamefont {Ripoll}, \citenamefont {Sagues}, \citenamefont
  {Doostmohammadi}, \citenamefont {Yeomans}, \citenamefont {Aranson},
  \citenamefont {Bechinger}, \citenamefont {Stark}, \citenamefont {Hemelrijk},
  \citenamefont {Nedelec}, \citenamefont {Sarkar}, \citenamefont {Aryaksama},
  \citenamefont {Lacroix}, \citenamefont {Duclos}, \citenamefont {Yashunsky},
  \citenamefont {Silberzan}, \citenamefont {Arroyo},\ and\ \citenamefont
  {Kale}}]{gompper2020}%
  \BibitemOpen
  \bibfield  {author} {\bibinfo {author} {\bibfnamefont {G.}~\bibnamefont
  {Gompper}}, \bibinfo {author} {\bibfnamefont {R.~G.}\ \bibnamefont
  {Winkler}}, \bibinfo {author} {\bibfnamefont {T.}~\bibnamefont {Speck}},
  \bibinfo {author} {\bibfnamefont {A.}~\bibnamefont {Solon}}, \bibinfo
  {author} {\bibfnamefont {C.}~\bibnamefont {Nardini}}, \bibinfo {author}
  {\bibfnamefont {F.}~\bibnamefont {Peruani}}, \bibinfo {author} {\bibfnamefont
  {H.}~\bibnamefont {L{\"o}wen}}, \bibinfo {author} {\bibfnamefont
  {R.}~\bibnamefont {Golestanian}}, \bibinfo {author} {\bibfnamefont {U.~B.}\
  \bibnamefont {Kaupp}}, \bibinfo {author} {\bibfnamefont {L.}~\bibnamefont
  {Alvarez}}, \bibinfo {author} {\bibfnamefont {T.}~\bibnamefont {Ki{\o}rboe}},
  \bibinfo {author} {\bibfnamefont {E.}~\bibnamefont {Lauga}}, \bibinfo
  {author} {\bibfnamefont {W.~C.~K.}\ \bibnamefont {Poon}}, \bibinfo {author}
  {\bibfnamefont {A.}~\bibnamefont {DeSimone}}, \bibinfo {author}
  {\bibfnamefont {S.}~\bibnamefont {Mui{\~{n}}os-Landin}}, \bibinfo {author}
  {\bibfnamefont {A.}~\bibnamefont {Fischer}}, \bibinfo {author} {\bibfnamefont
  {N.~A.}\ \bibnamefont {S{\"o}ker}}, \bibinfo {author} {\bibfnamefont
  {F.}~\bibnamefont {Cichos}}, \bibinfo {author} {\bibfnamefont
  {R.}~\bibnamefont {Kapral}}, \bibinfo {author} {\bibfnamefont
  {P.}~\bibnamefont {Gaspard}}, \bibinfo {author} {\bibfnamefont
  {M.}~\bibnamefont {Ripoll}}, \bibinfo {author} {\bibfnamefont
  {F.}~\bibnamefont {Sagues}}, \bibinfo {author} {\bibfnamefont
  {A.}~\bibnamefont {Doostmohammadi}}, \bibinfo {author} {\bibfnamefont
  {J.~M.}\ \bibnamefont {Yeomans}}, \bibinfo {author} {\bibfnamefont {I.~S.}\
  \bibnamefont {Aranson}}, \bibinfo {author} {\bibfnamefont {C.}~\bibnamefont
  {Bechinger}}, \bibinfo {author} {\bibfnamefont {H.}~\bibnamefont {Stark}},
  \bibinfo {author} {\bibfnamefont {C.~K.}\ \bibnamefont {Hemelrijk}}, \bibinfo
  {author} {\bibfnamefont {F.~J.}\ \bibnamefont {Nedelec}}, \bibinfo {author}
  {\bibfnamefont {T.}~\bibnamefont {Sarkar}}, \bibinfo {author} {\bibfnamefont
  {T.}~\bibnamefont {Aryaksama}}, \bibinfo {author} {\bibfnamefont
  {M.}~\bibnamefont {Lacroix}}, \bibinfo {author} {\bibfnamefont
  {G.}~\bibnamefont {Duclos}}, \bibinfo {author} {\bibfnamefont
  {V.}~\bibnamefont {Yashunsky}}, \bibinfo {author} {\bibfnamefont
  {P.}~\bibnamefont {Silberzan}}, \bibinfo {author} {\bibfnamefont
  {M.}~\bibnamefont {Arroyo}},\ and\ \bibinfo {author} {\bibfnamefont
  {S.}~\bibnamefont {Kale}},\ }\href {https://doi.org/10.1088/1361-648x/ab6348}
  {\bibfield  {journal} {\bibinfo  {journal} {J. Phys.: Condens. Matter}\
  }\textbf {\bibinfo {volume} {32}},\ \bibinfo {pages} {193001} (\bibinfo
  {year} {2020})}\BibitemShut {NoStop}%
\bibitem [{\citenamefont {Prost}\ \emph {et~al.}(2015)\citenamefont {Prost},
  \citenamefont {J{\"u}licher},\ and\ \citenamefont {Joanny}}]{Prost2015}%
  \BibitemOpen
  \bibfield  {author} {\bibinfo {author} {\bibfnamefont {J.}~\bibnamefont
  {Prost}}, \bibinfo {author} {\bibfnamefont {F.}~\bibnamefont
  {J{\"u}licher}},\ and\ \bibinfo {author} {\bibfnamefont {J.-F.}\ \bibnamefont
  {Joanny}},\ }\href {https://doi.org/10.1038/nphys3224} {\bibfield  {journal}
  {\bibinfo  {journal} {Nature Physics}\ }\textbf {\bibinfo {volume} {11}},\
  \bibinfo {pages} {111} (\bibinfo {year} {2015})}\BibitemShut {NoStop}%
\bibitem [{\citenamefont {Toner}\ and\ \citenamefont
  {Tu}(1998)}]{toner1998flocks}%
  \BibitemOpen
  \bibfield  {author} {\bibinfo {author} {\bibfnamefont {J.}~\bibnamefont
  {Toner}}\ and\ \bibinfo {author} {\bibfnamefont {Y.}~\bibnamefont {Tu}},\
  }\href {https://doi.org/10.1103/PhysRevE.58.4828} {\bibfield  {journal}
  {\bibinfo  {journal} {Phys. Rev. E}\ }\textbf {\bibinfo {volume} {58}},\
  \bibinfo {pages} {4828} (\bibinfo {year} {1998})}\BibitemShut {NoStop}%
\bibitem [{\citenamefont {Marchetti}\ \emph {et~al.}(2013)\citenamefont
  {Marchetti}, \citenamefont {Joanny}, \citenamefont {Ramaswamy}, \citenamefont
  {Liverpool}, \citenamefont {Prost}, \citenamefont {Rao},\ and\ \citenamefont
  {Simha}}]{marchetti2013hydrodynamics}%
  \BibitemOpen
  \bibfield  {author} {\bibinfo {author} {\bibfnamefont {M.~C.}\ \bibnamefont
  {Marchetti}}, \bibinfo {author} {\bibfnamefont {J.~F.}\ \bibnamefont
  {Joanny}}, \bibinfo {author} {\bibfnamefont {S.}~\bibnamefont {Ramaswamy}},
  \bibinfo {author} {\bibfnamefont {T.~B.}\ \bibnamefont {Liverpool}}, \bibinfo
  {author} {\bibfnamefont {J.}~\bibnamefont {Prost}}, \bibinfo {author}
  {\bibfnamefont {M.}~\bibnamefont {Rao}},\ and\ \bibinfo {author}
  {\bibfnamefont {R.~A.}\ \bibnamefont {Simha}},\ }\href
  {https://doi.org/10.1103/RevModPhys.85.1143} {\bibfield  {journal} {\bibinfo
  {journal} {Rev. Mod. Phys.}\ }\textbf {\bibinfo {volume} {85}},\ \bibinfo
  {pages} {1143} (\bibinfo {year} {2013})}\BibitemShut {NoStop}%
\bibitem [{\citenamefont {Bardeen}\ \emph {et~al.}(1957)\citenamefont
  {Bardeen}, \citenamefont {Cooper},\ and\ \citenamefont
  {Schrieffer}}]{BCS1957}%
  \BibitemOpen
  \bibfield  {author} {\bibinfo {author} {\bibfnamefont {J.}~\bibnamefont
  {Bardeen}}, \bibinfo {author} {\bibfnamefont {L.~N.}\ \bibnamefont
  {Cooper}},\ and\ \bibinfo {author} {\bibfnamefont {J.~R.}\ \bibnamefont
  {Schrieffer}},\ }\href {https://doi.org/10.1103/PhysRev.108.1175} {\bibfield
  {journal} {\bibinfo  {journal} {Phys. Rev.}\ }\textbf {\bibinfo {volume}
  {108}},\ \bibinfo {pages} {1175} (\bibinfo {year} {1957})}\BibitemShut
  {NoStop}%
\bibitem [{\citenamefont {Lee}\ \emph {et~al.}(2006)\citenamefont {Lee},
  \citenamefont {Nagaosa},\ and\ \citenamefont {Wen}}]{LeeRMP2006}%
  \BibitemOpen
  \bibfield  {author} {\bibinfo {author} {\bibfnamefont {P.~A.}\ \bibnamefont
  {Lee}}, \bibinfo {author} {\bibfnamefont {N.}~\bibnamefont {Nagaosa}},\ and\
  \bibinfo {author} {\bibfnamefont {X.-G.}\ \bibnamefont {Wen}},\ }\href
  {https://doi.org/10.1103/RevModPhys.78.17} {\bibfield  {journal} {\bibinfo
  {journal} {Rev. Mod. Phys.}\ }\textbf {\bibinfo {volume} {78}},\ \bibinfo
  {pages} {17} (\bibinfo {year} {2006})}\BibitemShut {NoStop}%
\bibitem [{\citenamefont {Soto}\ and\ \citenamefont
  {Golestanian}(2014)}]{soto2014self}%
  \BibitemOpen
  \bibfield  {author} {\bibinfo {author} {\bibfnamefont {R.}~\bibnamefont
  {Soto}}\ and\ \bibinfo {author} {\bibfnamefont {R.}~\bibnamefont
  {Golestanian}},\ }\href {https://doi.org/10.1103/PhysRevLett.112.068301}
  {\bibfield  {journal} {\bibinfo  {journal} {Phys. Rev. Lett.}\ }\textbf
  {\bibinfo {volume} {112}},\ \bibinfo {pages} {068301} (\bibinfo {year}
  {2014})}\BibitemShut {NoStop}%
\bibitem [{\citenamefont {Golestanian}(2022)}]{Golestanian2019phoretic}%
  \BibitemOpen
  \bibfield  {author} {\bibinfo {author} {\bibfnamefont {R.}~\bibnamefont
  {Golestanian}},\ }in\ \href
  {https://doi.org/10.1093/oso/9780192858313.003.0008} {\emph {\bibinfo
  {booktitle} {{Active Matter and Nonequilibrium Statistical Physics: Lecture
  Notes of the Les Houches Summer School: Volume 112, September 2018}}}}\
  (\bibinfo  {publisher} {Oxford University Press},\ \bibinfo {year}
  {2022})\BibitemShut {NoStop}%
\bibitem [{\citenamefont {Niu}\ \emph {et~al.}(2018)\citenamefont {Niu},
  \citenamefont {Fischer}, \citenamefont {Palberg},\ and\ \citenamefont
  {Speck}}]{niu2018dynamics}%
  \BibitemOpen
  \bibfield  {author} {\bibinfo {author} {\bibfnamefont {R.}~\bibnamefont
  {Niu}}, \bibinfo {author} {\bibfnamefont {A.}~\bibnamefont {Fischer}},
  \bibinfo {author} {\bibfnamefont {T.}~\bibnamefont {Palberg}},\ and\ \bibinfo
  {author} {\bibfnamefont {T.}~\bibnamefont {Speck}},\ }\href
  {https://doi.org/10.1021/acsnano.8b04221} {\bibfield  {journal} {\bibinfo
  {journal} {ACS Nano}\ }\textbf {\bibinfo {volume} {12}},\ \bibinfo {pages}
  {10932–10938} (\bibinfo {year} {2018})}\BibitemShut {NoStop}%
\bibitem [{\citenamefont {Meredith}\ \emph {et~al.}(2020)\citenamefont
  {Meredith}, \citenamefont {Moerman}, \citenamefont {Groenewold},
  \citenamefont {Chiu}, \citenamefont {Kegel}, \citenamefont {van Blaaderen},\
  and\ \citenamefont {Zarzar}}]{meredith2020predator}%
  \BibitemOpen
  \bibfield  {author} {\bibinfo {author} {\bibfnamefont {C.~H.}\ \bibnamefont
  {Meredith}}, \bibinfo {author} {\bibfnamefont {P.~G.}\ \bibnamefont
  {Moerman}}, \bibinfo {author} {\bibfnamefont {J.}~\bibnamefont {Groenewold}},
  \bibinfo {author} {\bibfnamefont {Y.-J.}\ \bibnamefont {Chiu}}, \bibinfo
  {author} {\bibfnamefont {W.~K.}\ \bibnamefont {Kegel}}, \bibinfo {author}
  {\bibfnamefont {A.}~\bibnamefont {van Blaaderen}},\ and\ \bibinfo {author}
  {\bibfnamefont {L.~D.}\ \bibnamefont {Zarzar}},\ }\href
  {https://doi.org/10.1038/s41557-020-00575-0} {\bibfield  {journal} {\bibinfo
  {journal} {Nature Chemistry}\ }\textbf {\bibinfo {volume} {12}},\ \bibinfo
  {pages} {1136–1142} (\bibinfo {year} {2020})}\BibitemShut {NoStop}%
\bibitem [{\citenamefont {Ivlev}\ \emph {et~al.}(2015)\citenamefont {Ivlev},
  \citenamefont {Bartnick}, \citenamefont {Heinen}, \citenamefont {Du},
  \citenamefont {Nosenko},\ and\ \citenamefont
  {L\"owen}}]{ivlev2015statistical}%
  \BibitemOpen
  \bibfield  {author} {\bibinfo {author} {\bibfnamefont {A.~V.}\ \bibnamefont
  {Ivlev}}, \bibinfo {author} {\bibfnamefont {J.}~\bibnamefont {Bartnick}},
  \bibinfo {author} {\bibfnamefont {M.}~\bibnamefont {Heinen}}, \bibinfo
  {author} {\bibfnamefont {C.-R.}\ \bibnamefont {Du}}, \bibinfo {author}
  {\bibfnamefont {V.}~\bibnamefont {Nosenko}},\ and\ \bibinfo {author}
  {\bibfnamefont {H.}~\bibnamefont {L\"owen}},\ }\href
  {https://doi.org/10.1103/PhysRevX.5.011035} {\bibfield  {journal} {\bibinfo
  {journal} {Phys. Rev. X}\ }\textbf {\bibinfo {volume} {5}},\ \bibinfo {pages}
  {011035} (\bibinfo {year} {2015})}\BibitemShut {NoStop}%
\bibitem [{\citenamefont {Najafi}\ and\ \citenamefont
  {Golestanian}(2004)}]{3SS}%
  \BibitemOpen
  \bibfield  {author} {\bibinfo {author} {\bibfnamefont {A.}~\bibnamefont
  {Najafi}}\ and\ \bibinfo {author} {\bibfnamefont {R.}~\bibnamefont
  {Golestanian}},\ }\href {https://doi.org/10.1103/PhysRevE.69.062901}
  {\bibfield  {journal} {\bibinfo  {journal} {Phys. Rev. E}\ }\textbf {\bibinfo
  {volume} {69}},\ \bibinfo {pages} {062901} (\bibinfo {year}
  {2004})}\BibitemShut {NoStop}%
\bibitem [{\citenamefont {Uchida}\ and\ \citenamefont
  {Golestanian}(2010{\natexlab{a}})}]{uchida2010synchronization}%
  \BibitemOpen
  \bibfield  {author} {\bibinfo {author} {\bibfnamefont {N.}~\bibnamefont
  {Uchida}}\ and\ \bibinfo {author} {\bibfnamefont {R.}~\bibnamefont
  {Golestanian}},\ }\href {https://doi.org/10.1103/PhysRevLett.104.178103}
  {\bibfield  {journal} {\bibinfo  {journal} {Phys. Rev. Lett.}\ }\textbf
  {\bibinfo {volume} {104}},\ \bibinfo {pages} {178103} (\bibinfo {year}
  {2010}{\natexlab{a}})}\BibitemShut {NoStop}%
\bibitem [{\citenamefont {Saha}\ \emph {et~al.}(2019)\citenamefont {Saha},
  \citenamefont {Ramaswamy},\ and\ \citenamefont {Golestanian}}]{Saha2019}%
  \BibitemOpen
  \bibfield  {author} {\bibinfo {author} {\bibfnamefont {S.}~\bibnamefont
  {Saha}}, \bibinfo {author} {\bibfnamefont {S.}~\bibnamefont {Ramaswamy}},\
  and\ \bibinfo {author} {\bibfnamefont {R.}~\bibnamefont {Golestanian}},\
  }\href {https://doi.org/10.1088/1367-2630/ab20fd} {\bibfield  {journal}
  {\bibinfo  {journal} {New Journal of Physics}\ }\textbf {\bibinfo {volume}
  {21}},\ \bibinfo {pages} {063006} (\bibinfo {year} {2019})}\BibitemShut
  {NoStop}%
\bibitem [{\citenamefont {Dadhichi}\ \emph {et~al.}(2020)\citenamefont
  {Dadhichi}, \citenamefont {Kethapelli}, \citenamefont {Chajwa}, \citenamefont
  {Ramaswamy},\ and\ \citenamefont {Maitra}}]{Dadhichi_PhysRevE.101.052601}%
  \BibitemOpen
  \bibfield  {author} {\bibinfo {author} {\bibfnamefont {L.~P.}\ \bibnamefont
  {Dadhichi}}, \bibinfo {author} {\bibfnamefont {J.}~\bibnamefont
  {Kethapelli}}, \bibinfo {author} {\bibfnamefont {R.}~\bibnamefont {Chajwa}},
  \bibinfo {author} {\bibfnamefont {S.}~\bibnamefont {Ramaswamy}},\ and\
  \bibinfo {author} {\bibfnamefont {A.}~\bibnamefont {Maitra}},\ }\href
  {https://doi.org/10.1103/PhysRevE.101.052601} {\bibfield  {journal} {\bibinfo
   {journal} {Phys. Rev. E}\ }\textbf {\bibinfo {volume} {101}},\ \bibinfo
  {pages} {052601} (\bibinfo {year} {2020})}\BibitemShut {NoStop}%
\bibitem [{\citenamefont {Fruchart}\ \emph {et~al.}(2021)\citenamefont
  {Fruchart}, \citenamefont {Hanai}, \citenamefont {Littlewood},\ and\
  \citenamefont {Vitelli}}]{fruchart2021non}%
  \BibitemOpen
  \bibfield  {author} {\bibinfo {author} {\bibfnamefont {M.}~\bibnamefont
  {Fruchart}}, \bibinfo {author} {\bibfnamefont {R.}~\bibnamefont {Hanai}},
  \bibinfo {author} {\bibfnamefont {P.~B.}\ \bibnamefont {Littlewood}},\ and\
  \bibinfo {author} {\bibfnamefont {V.}~\bibnamefont {Vitelli}},\ }\href
  {https://doi.org/10.1038/s41586-021-03375-9} {\bibfield  {journal} {\bibinfo
  {journal} {Nature}\ }\textbf {\bibinfo {volume} {592}},\ \bibinfo {pages}
  {363–369} (\bibinfo {year} {2021})}\BibitemShut {NoStop}%
\bibitem [{\citenamefont {Kreienkamp}\ and\ \citenamefont
  {Klapp}(2022)}]{kreienkamp2022clustering}%
  \BibitemOpen
  \bibfield  {author} {\bibinfo {author} {\bibfnamefont {K.~L.}\ \bibnamefont
  {Kreienkamp}}\ and\ \bibinfo {author} {\bibfnamefont {S.~H.~L.}\ \bibnamefont
  {Klapp}},\ }\href {https://doi.org/10.1088/1367-2630/ac9cc3} {\bibfield
  {journal} {\bibinfo  {journal} {New Journal of Physics}\ }\textbf {\bibinfo
  {volume} {24}},\ \bibinfo {pages} {123009} (\bibinfo {year}
  {2022})}\BibitemShut {NoStop}%
\bibitem [{\citenamefont {Loos}\ \emph {et~al.}(2023)\citenamefont {Loos},
  \citenamefont {Klapp},\ and\ \citenamefont {Martynec}}]{Loos2023}%
  \BibitemOpen
  \bibfield  {author} {\bibinfo {author} {\bibfnamefont {S.~A.~M.}\
  \bibnamefont {Loos}}, \bibinfo {author} {\bibfnamefont {S.~H.~L.}\
  \bibnamefont {Klapp}},\ and\ \bibinfo {author} {\bibfnamefont
  {T.}~\bibnamefont {Martynec}},\ }\href
  {https://doi.org/10.1103/PhysRevLett.130.198301} {\bibfield  {journal}
  {\bibinfo  {journal} {Phys. Rev. Lett.}\ }\textbf {\bibinfo {volume} {130}},\
  \bibinfo {pages} {198301} (\bibinfo {year} {2023})}\BibitemShut {NoStop}%
\bibitem [{\citenamefont {Agudo-Canalejo}\ and\ \citenamefont
  {Golestanian}(2019)}]{agudo2019active}%
  \BibitemOpen
  \bibfield  {author} {\bibinfo {author} {\bibfnamefont {J.}~\bibnamefont
  {Agudo-Canalejo}}\ and\ \bibinfo {author} {\bibfnamefont {R.}~\bibnamefont
  {Golestanian}},\ }\href {https://doi.org/10.1103/PhysRevLett.123.018101}
  {\bibfield  {journal} {\bibinfo  {journal} {Phys. Rev. Lett.}\ }\textbf
  {\bibinfo {volume} {123}},\ \bibinfo {pages} {018101} (\bibinfo {year}
  {2019})}\BibitemShut {NoStop}%
\bibitem [{\citenamefont {Saha}\ \emph {et~al.}(2020)\citenamefont {Saha},
  \citenamefont {Agudo-Canalejo},\ and\ \citenamefont
  {Golestanian}}]{saha2020scalar}%
  \BibitemOpen
  \bibfield  {author} {\bibinfo {author} {\bibfnamefont {S.}~\bibnamefont
  {Saha}}, \bibinfo {author} {\bibfnamefont {J.}~\bibnamefont
  {Agudo-Canalejo}},\ and\ \bibinfo {author} {\bibfnamefont {R.}~\bibnamefont
  {Golestanian}},\ }\href {https://doi.org/10.1103/PhysRevX.10.041009}
  {\bibfield  {journal} {\bibinfo  {journal} {Phys. Rev. X}\ }\textbf {\bibinfo
  {volume} {10}},\ \bibinfo {pages} {041009} (\bibinfo {year}
  {2020})}\BibitemShut {NoStop}%
\bibitem [{\citenamefont {You}\ \emph {et~al.}(2020)\citenamefont {You},
  \citenamefont {Baskaran},\ and\ \citenamefont
  {Marchetti}}]{you2020nonreciprocity}%
  \BibitemOpen
  \bibfield  {author} {\bibinfo {author} {\bibfnamefont {Z.}~\bibnamefont
  {You}}, \bibinfo {author} {\bibfnamefont {A.}~\bibnamefont {Baskaran}},\ and\
  \bibinfo {author} {\bibfnamefont {M.~C.}\ \bibnamefont {Marchetti}},\ }\href
  {https://doi.org/10.1073/pnas.2010318117} {\bibfield  {journal} {\bibinfo
  {journal} {Proceedings of the National Academy of Sciences}\ }\textbf
  {\bibinfo {volume} {117}},\ \bibinfo {pages} {19767–19772} (\bibinfo {year}
  {2020})}\BibitemShut {NoStop}%
\bibitem [{\citenamefont {Frohoff-H\"ulsmann}\ \emph
  {et~al.}(2021)\citenamefont {Frohoff-H\"ulsmann}, \citenamefont {Wrembel},\
  and\ \citenamefont {Thiele}}]{frohoff2021suppression}%
  \BibitemOpen
  \bibfield  {author} {\bibinfo {author} {\bibfnamefont {T.}~\bibnamefont
  {Frohoff-H\"ulsmann}}, \bibinfo {author} {\bibfnamefont {J.}~\bibnamefont
  {Wrembel}},\ and\ \bibinfo {author} {\bibfnamefont {U.}~\bibnamefont
  {Thiele}},\ }\href {https://doi.org/10.1103/PhysRevE.103.042602} {\bibfield
  {journal} {\bibinfo  {journal} {Phys. Rev. E}\ }\textbf {\bibinfo {volume}
  {103}},\ \bibinfo {pages} {042602} (\bibinfo {year} {2021})}\BibitemShut
  {NoStop}%
\bibitem [{\citenamefont {Dinelli}\ \emph {et~al.}(2023)\citenamefont
  {Dinelli}, \citenamefont {O’Byrne}, \citenamefont {Curatolo}, \citenamefont
  {Zhao}, \citenamefont {Sollich},\ and\ \citenamefont
  {Tailleur}}]{dinelli2023non}%
  \BibitemOpen
  \bibfield  {author} {\bibinfo {author} {\bibfnamefont {A.}~\bibnamefont
  {Dinelli}}, \bibinfo {author} {\bibfnamefont {J.}~\bibnamefont {O’Byrne}},
  \bibinfo {author} {\bibfnamefont {A.}~\bibnamefont {Curatolo}}, \bibinfo
  {author} {\bibfnamefont {Y.}~\bibnamefont {Zhao}}, \bibinfo {author}
  {\bibfnamefont {P.}~\bibnamefont {Sollich}},\ and\ \bibinfo {author}
  {\bibfnamefont {J.}~\bibnamefont {Tailleur}},\ }\href@noop {} {\bibfield
  {journal} {\bibinfo  {journal} {Nature Communications}\ }\textbf {\bibinfo
  {volume} {14}},\ \bibinfo {pages} {7035} (\bibinfo {year}
  {2023})}\BibitemShut {NoStop}%
\bibitem [{\citenamefont {Weis}\ \emph {et~al.}(2022)\citenamefont {Weis},
  \citenamefont {Fruchart}, \citenamefont {Hanai}, \citenamefont {Kawagoe},
  \citenamefont {Littlewood},\ and\ \citenamefont
  {Vitelli}}]{weis2022coalescence}%
  \BibitemOpen
  \bibfield  {author} {\bibinfo {author} {\bibfnamefont {C.}~\bibnamefont
  {Weis}}, \bibinfo {author} {\bibfnamefont {M.}~\bibnamefont {Fruchart}},
  \bibinfo {author} {\bibfnamefont {R.}~\bibnamefont {Hanai}}, \bibinfo
  {author} {\bibfnamefont {K.}~\bibnamefont {Kawagoe}}, \bibinfo {author}
  {\bibfnamefont {P.~B.}\ \bibnamefont {Littlewood}},\ and\ \bibinfo {author}
  {\bibfnamefont {V.}~\bibnamefont {Vitelli}},\ }\href
  {https://doi.org/10.48550/ARXIV.2207.11667} {\bibinfo {title} {Exceptional
  points in nonlinear and stochastic dynamics}} (\bibinfo {year}
  {2022})\BibitemShut {NoStop}%
\bibitem [{\citenamefont {Saha}\ and\ \citenamefont
  {Golestanian}(2022)}]{saha2022effervescent}%
  \BibitemOpen
  \bibfield  {author} {\bibinfo {author} {\bibfnamefont {S.}~\bibnamefont
  {Saha}}\ and\ \bibinfo {author} {\bibfnamefont {R.}~\bibnamefont
  {Golestanian}},\ }\href {https://doi.org/10.48550/ARXIV.2208.14985} {\bibinfo
  {title} {Effervescent waves in a binary mixture with non-reciprocal
  couplings}} (\bibinfo {year} {2022})\BibitemShut {NoStop}%
\bibitem [{\citenamefont {Soto}\ and\ \citenamefont
  {Golestanian}(2015)}]{soto2015self}%
  \BibitemOpen
  \bibfield  {author} {\bibinfo {author} {\bibfnamefont {R.}~\bibnamefont
  {Soto}}\ and\ \bibinfo {author} {\bibfnamefont {R.}~\bibnamefont
  {Golestanian}},\ }\href {https://doi.org/10.1103/PhysRevE.91.052304}
  {\bibfield  {journal} {\bibinfo  {journal} {Phys. Rev. E}\ }\textbf {\bibinfo
  {volume} {91}},\ \bibinfo {pages} {052304} (\bibinfo {year}
  {2015})}\BibitemShut {NoStop}%
\bibitem [{\citenamefont {Osat}\ and\ \citenamefont
  {Golestanian}(2022)}]{osat2023non}%
  \BibitemOpen
  \bibfield  {author} {\bibinfo {author} {\bibfnamefont {S.}~\bibnamefont
  {Osat}}\ and\ \bibinfo {author} {\bibfnamefont {R.}~\bibnamefont
  {Golestanian}},\ }\href {https://doi.org/10.1038/s41565-022-01258-2}
  {\bibfield  {journal} {\bibinfo  {journal} {Nature Nanotechnology}\ }\textbf
  {\bibinfo {volume} {18}},\ \bibinfo {pages} {79–85} (\bibinfo {year}
  {2022})}\BibitemShut {NoStop}%
\bibitem [{\citenamefont {Ouazan-Reboul}\ \emph
  {et~al.}(2023{\natexlab{a}})\citenamefont {Ouazan-Reboul}, \citenamefont
  {Agudo-Canalejo},\ and\ \citenamefont {Golestanian}}]{OuazanReboul2023-I}%
  \BibitemOpen
  \bibfield  {author} {\bibinfo {author} {\bibfnamefont {V.}~\bibnamefont
  {Ouazan-Reboul}}, \bibinfo {author} {\bibfnamefont {J.}~\bibnamefont
  {Agudo-Canalejo}},\ and\ \bibinfo {author} {\bibfnamefont {R.}~\bibnamefont
  {Golestanian}},\ }\bibfield  {journal} {\bibinfo  {journal} {Nature
  Communications}\ }\textbf {\bibinfo {volume} {14}},\ \href
  {https://doi.org/10.1038/s41467-023-40241-w} {10.1038/s41467-023-40241-w}
  (\bibinfo {year} {2023}{\natexlab{a}})\BibitemShut {NoStop}%
\bibitem [{\citenamefont {Ouazan-Reboul}\ \emph
  {et~al.}(2023{\natexlab{b}})\citenamefont {Ouazan-Reboul}, \citenamefont
  {Golestanian},\ and\ \citenamefont {Agudo-Canalejo}}]{OuazanReboul2023-II}%
  \BibitemOpen
  \bibfield  {author} {\bibinfo {author} {\bibfnamefont {V.}~\bibnamefont
  {Ouazan-Reboul}}, \bibinfo {author} {\bibfnamefont {R.}~\bibnamefont
  {Golestanian}},\ and\ \bibinfo {author} {\bibfnamefont {J.}~\bibnamefont
  {Agudo-Canalejo}},\ }\href {https://doi.org/10.1103/PhysRevLett.131.128301}
  {\bibfield  {journal} {\bibinfo  {journal} {Phys. Rev. Lett.}\ }\textbf
  {\bibinfo {volume} {131}},\ \bibinfo {pages} {128301} (\bibinfo {year}
  {2023}{\natexlab{b}})}\BibitemShut {NoStop}%
\bibitem [{\citenamefont {Ouazan-Reboul}\ \emph
  {et~al.}(2023{\natexlab{c}})\citenamefont {Ouazan-Reboul}, \citenamefont
  {Golestanian},\ and\ \citenamefont {Agudo-Canalejo}}]{OuazanReboul2023-III}%
  \BibitemOpen
  \bibfield  {author} {\bibinfo {author} {\bibfnamefont {V.}~\bibnamefont
  {Ouazan-Reboul}}, \bibinfo {author} {\bibfnamefont {R.}~\bibnamefont
  {Golestanian}},\ and\ \bibinfo {author} {\bibfnamefont {J.}~\bibnamefont
  {Agudo-Canalejo}},\ }\href {https://doi.org/10.1088/1367-2630/acfdc2}
  {\bibfield  {journal} {\bibinfo  {journal} {New Journal of Physics}\ }\textbf
  {\bibinfo {volume} {25}},\ \bibinfo {pages} {103013} (\bibinfo {year}
  {2023}{\natexlab{c}})}\BibitemShut {NoStop}%
\bibitem [{\citenamefont {Toner}\ and\ \citenamefont
  {Tu}(1995)}]{toner1995long}%
  \BibitemOpen
  \bibfield  {author} {\bibinfo {author} {\bibfnamefont {J.}~\bibnamefont
  {Toner}}\ and\ \bibinfo {author} {\bibfnamefont {Y.}~\bibnamefont {Tu}},\
  }\href {https://doi.org/10.1103/PhysRevLett.75.4326} {\bibfield  {journal}
  {\bibinfo  {journal} {Phys. Rev. Lett.}\ }\textbf {\bibinfo {volume} {75}},\
  \bibinfo {pages} {4326} (\bibinfo {year} {1995})}\BibitemShut {NoStop}%
\bibitem [{\citenamefont {Toner}(2012{\natexlab{a}})}]{Toner2012PRL}%
  \BibitemOpen
  \bibfield  {author} {\bibinfo {author} {\bibfnamefont {J.}~\bibnamefont
  {Toner}},\ }\href {https://doi.org/10.1103/PhysRevLett.108.088102} {\bibfield
   {journal} {\bibinfo  {journal} {Phys. Rev. Lett.}\ }\textbf {\bibinfo
  {volume} {108}},\ \bibinfo {pages} {088102} (\bibinfo {year}
  {2012}{\natexlab{a}})}\BibitemShut {NoStop}%
\bibitem [{\citenamefont {Toner}(2012{\natexlab{b}})}]{Toner2012PRE}%
  \BibitemOpen
  \bibfield  {author} {\bibinfo {author} {\bibfnamefont {J.}~\bibnamefont
  {Toner}},\ }\href {https://doi.org/10.1103/PhysRevE.86.031918} {\bibfield
  {journal} {\bibinfo  {journal} {Phys. Rev. E}\ }\textbf {\bibinfo {volume}
  {86}},\ \bibinfo {pages} {031918} (\bibinfo {year}
  {2012}{\natexlab{b}})}\BibitemShut {NoStop}%
\bibitem [{\citenamefont {Kardar}\ \emph {et~al.}(1986)\citenamefont {Kardar},
  \citenamefont {Parisi},\ and\ \citenamefont {Zhang}}]{kardar1986dynamic}%
  \BibitemOpen
  \bibfield  {author} {\bibinfo {author} {\bibfnamefont {M.}~\bibnamefont
  {Kardar}}, \bibinfo {author} {\bibfnamefont {G.}~\bibnamefont {Parisi}},\
  and\ \bibinfo {author} {\bibfnamefont {Y.-C.}\ \bibnamefont {Zhang}},\ }\href
  {https://doi.org/10.1103/PhysRevLett.56.889} {\bibfield  {journal} {\bibinfo
  {journal} {Phys. Rev. Lett.}\ }\textbf {\bibinfo {volume} {56}},\ \bibinfo
  {pages} {889} (\bibinfo {year} {1986})}\BibitemShut {NoStop}%
\bibitem [{\citenamefont {Forster}\ \emph {et~al.}(1977)\citenamefont
  {Forster}, \citenamefont {Nelson},\ and\ \citenamefont
  {Stephen}}]{forster1977large}%
  \BibitemOpen
  \bibfield  {author} {\bibinfo {author} {\bibfnamefont {D.}~\bibnamefont
  {Forster}}, \bibinfo {author} {\bibfnamefont {D.~R.}\ \bibnamefont
  {Nelson}},\ and\ \bibinfo {author} {\bibfnamefont {M.~J.}\ \bibnamefont
  {Stephen}},\ }\href {https://doi.org/10.1103/PhysRevA.16.732} {\bibfield
  {journal} {\bibinfo  {journal} {Phys. Rev. A}\ }\textbf {\bibinfo {volume}
  {16}},\ \bibinfo {pages} {732} (\bibinfo {year} {1977})}\BibitemShut
  {NoStop}%
\bibitem [{\citenamefont {Frey}\ and\ \citenamefont
  {T\"{a}uber}(1994)}]{Frey1994}%
  \BibitemOpen
  \bibfield  {author} {\bibinfo {author} {\bibfnamefont {E.}~\bibnamefont
  {Frey}}\ and\ \bibinfo {author} {\bibfnamefont {U.~C.}\ \bibnamefont
  {T\"{a}uber}},\ }\href {https://doi.org/10.1103/physreve.50.1024} {\bibfield
  {journal} {\bibinfo  {journal} {Physical Review E}\ }\textbf {\bibinfo
  {volume} {50}},\ \bibinfo {pages} {1024–1045} (\bibinfo {year}
  {1994})}\BibitemShut {NoStop}%
\bibitem [{\citenamefont {Mermin}\ and\ \citenamefont
  {Wagner}(1966)}]{mermin1966absence}%
  \BibitemOpen
  \bibfield  {author} {\bibinfo {author} {\bibfnamefont {N.~D.}\ \bibnamefont
  {Mermin}}\ and\ \bibinfo {author} {\bibfnamefont {H.}~\bibnamefont
  {Wagner}},\ }\href {https://doi.org/10.1103/PhysRevLett.17.1133} {\bibfield
  {journal} {\bibinfo  {journal} {Phys. Rev. Lett.}\ }\textbf {\bibinfo
  {volume} {17}},\ \bibinfo {pages} {1133} (\bibinfo {year}
  {1966})}\BibitemShut {NoStop}%
\bibitem [{\citenamefont {Frohoff-H\"{u}lsmann}\ and\ \citenamefont
  {Thiele}(2021)}]{frohoff2021localized}%
  \BibitemOpen
  \bibfield  {author} {\bibinfo {author} {\bibfnamefont {T.}~\bibnamefont
  {Frohoff-H\"{u}lsmann}}\ and\ \bibinfo {author} {\bibfnamefont
  {U.}~\bibnamefont {Thiele}},\ }\href {https://doi.org/10.1093/imamat/hxab026}
  {\bibfield  {journal} {\bibinfo  {journal} {IMA Journal of Applied
  Mathematics}\ }\textbf {\bibinfo {volume} {86}},\ \bibinfo {pages}
  {924–943} (\bibinfo {year} {2021})}\BibitemShut {NoStop}%
\bibitem [{\citenamefont {Rana}\ and\ \citenamefont
  {Golestanian}(2024)}]{rana2023defect}%
  \BibitemOpen
  \bibfield  {author} {\bibinfo {author} {\bibfnamefont {N.}~\bibnamefont
  {Rana}}\ and\ \bibinfo {author} {\bibfnamefont {R.}~\bibnamefont
  {Golestanian}},\ }\href {https://doi.org/10.1103/PhysRevLett.133.078301}
  {\bibfield  {journal} {\bibinfo  {journal} {Phys. Rev. Lett.}\ }\textbf
  {\bibinfo {volume} {133}},\ \bibinfo {pages} {078301} (\bibinfo {year}
  {2024})}\BibitemShut {NoStop}%
\bibitem [{\citenamefont {Aranson}\ and\ \citenamefont
  {Kramer}(2002)}]{aranson2002world}%
  \BibitemOpen
  \bibfield  {author} {\bibinfo {author} {\bibfnamefont {I.~S.}\ \bibnamefont
  {Aranson}}\ and\ \bibinfo {author} {\bibfnamefont {L.}~\bibnamefont
  {Kramer}},\ }\href {https://doi.org/10.1103/RevModPhys.74.99} {\bibfield
  {journal} {\bibinfo  {journal} {Rev. Mod. Phys.}\ }\textbf {\bibinfo {volume}
  {74}},\ \bibinfo {pages} {99} (\bibinfo {year} {2002})}\BibitemShut {NoStop}%
\bibitem [{\citenamefont {Hohenberg}\ and\ \citenamefont
  {Halperin}(1977)}]{hohenberg1977theory}%
  \BibitemOpen
  \bibfield  {author} {\bibinfo {author} {\bibfnamefont {P.~C.}\ \bibnamefont
  {Hohenberg}}\ and\ \bibinfo {author} {\bibfnamefont {B.~I.}\ \bibnamefont
  {Halperin}},\ }\href {https://doi.org/10.1103/RevModPhys.49.435} {\bibfield
  {journal} {\bibinfo  {journal} {Rev. Mod. Phys.}\ }\textbf {\bibinfo {volume}
  {49}},\ \bibinfo {pages} {435} (\bibinfo {year} {1977})}\BibitemShut
  {NoStop}%
\bibitem [{\citenamefont {Medina}\ \emph {et~al.}(1989)\citenamefont {Medina},
  \citenamefont {Hwa}, \citenamefont {Kardar},\ and\ \citenamefont
  {Zhang}}]{medina1989burgers}%
  \BibitemOpen
  \bibfield  {author} {\bibinfo {author} {\bibfnamefont {E.}~\bibnamefont
  {Medina}}, \bibinfo {author} {\bibfnamefont {T.}~\bibnamefont {Hwa}},
  \bibinfo {author} {\bibfnamefont {M.}~\bibnamefont {Kardar}},\ and\ \bibinfo
  {author} {\bibfnamefont {Y.-C.}\ \bibnamefont {Zhang}},\ }\href
  {https://doi.org/10.1103/PhysRevA.39.3053} {\bibfield  {journal} {\bibinfo
  {journal} {Phys. Rev. A}\ }\textbf {\bibinfo {volume} {39}},\ \bibinfo
  {pages} {3053} (\bibinfo {year} {1989})}\BibitemShut {NoStop}%
\bibitem [{\citenamefont {Canet}\ \emph {et~al.}(2011)\citenamefont {Canet},
  \citenamefont {Chat{\'e}}, \citenamefont {Delamotte},\ and\ \citenamefont
  {Wschebor}}]{canet2011nonperturbative}%
  \BibitemOpen
  \bibfield  {author} {\bibinfo {author} {\bibfnamefont {L.}~\bibnamefont
  {Canet}}, \bibinfo {author} {\bibfnamefont {H.}~\bibnamefont {Chat{\'e}}},
  \bibinfo {author} {\bibfnamefont {B.}~\bibnamefont {Delamotte}},\ and\
  \bibinfo {author} {\bibfnamefont {N.}~\bibnamefont {Wschebor}},\ }\href@noop
  {} {\bibfield  {journal} {\bibinfo  {journal} {Physical Review E}\ }\textbf
  {\bibinfo {volume} {84}},\ \bibinfo {pages} {061128} (\bibinfo {year}
  {2011})}\BibitemShut {NoStop}%
\bibitem [{\citenamefont {Wiese}(1997)}]{wiese1997critical}%
  \BibitemOpen
  \bibfield  {author} {\bibinfo {author} {\bibfnamefont {K.~J.}\ \bibnamefont
  {Wiese}},\ }\href {https://doi.org/10.1103/PhysRevE.56.5013} {\bibfield
  {journal} {\bibinfo  {journal} {Phys. Rev. E}\ }\textbf {\bibinfo {volume}
  {56}},\ \bibinfo {pages} {5013} (\bibinfo {year} {1997})}\BibitemShut
  {NoStop}%
\bibitem [{\citenamefont {Oliveira}(2022)}]{KPZ-all-d}%
  \BibitemOpen
  \bibfield  {author} {\bibinfo {author} {\bibfnamefont {T.~J.}\ \bibnamefont
  {Oliveira}},\ }\href {https://doi.org/10.1103/PhysRevE.106.L062103}
  {\bibfield  {journal} {\bibinfo  {journal} {Phys. Rev. E}\ }\textbf {\bibinfo
  {volume} {106}},\ \bibinfo {pages} {L062103} (\bibinfo {year}
  {2022})}\BibitemShut {NoStop}%
\bibitem [{\citenamefont {Binney}\ \emph {et~al.}(1992)\citenamefont {Binney},
  \citenamefont {Dowrick}, \citenamefont {Fisher},\ and\ \citenamefont
  {Newman}}]{binney1992theory}%
  \BibitemOpen
  \bibfield  {author} {\bibinfo {author} {\bibfnamefont {J.~J.}\ \bibnamefont
  {Binney}}, \bibinfo {author} {\bibfnamefont {N.~J.}\ \bibnamefont {Dowrick}},
  \bibinfo {author} {\bibfnamefont {A.~J.}\ \bibnamefont {Fisher}},\ and\
  \bibinfo {author} {\bibfnamefont {M.~E.}\ \bibnamefont {Newman}},\
  }\href@noop {} {\emph {\bibinfo {title} {The theory of critical phenomena: an
  introduction to the renormalization group}}}\ (\bibinfo  {publisher} {Oxford
  University Press},\ \bibinfo {year} {1992})\BibitemShut {NoStop}%
\bibitem [{\citenamefont {Uchida}\ and\ \citenamefont
  {Golestanian}(2010{\natexlab{b}})}]{uchida2010synchronization2}%
  \BibitemOpen
  \bibfield  {author} {\bibinfo {author} {\bibfnamefont {N.}~\bibnamefont
  {Uchida}}\ and\ \bibinfo {author} {\bibfnamefont {R.}~\bibnamefont
  {Golestanian}},\ }\href {https://doi.org/10.1209/0295-5075/89/50011}
  {\bibfield  {journal} {\bibinfo  {journal} {Europhys. Lett.}\ }\textbf
  {\bibinfo {volume} {89}},\ \bibinfo {pages} {50011} (\bibinfo {year}
  {2010}{\natexlab{b}})}\BibitemShut {NoStop}%
\bibitem [{\citenamefont {Sarkar}\ \emph {et~al.}(2021)\citenamefont {Sarkar},
  \citenamefont {Basu},\ and\ \citenamefont {Toner}}]{sarkar2021swarming}%
  \BibitemOpen
  \bibfield  {author} {\bibinfo {author} {\bibfnamefont {N.}~\bibnamefont
  {Sarkar}}, \bibinfo {author} {\bibfnamefont {A.}~\bibnamefont {Basu}},\ and\
  \bibinfo {author} {\bibfnamefont {J.}~\bibnamefont {Toner}},\ }\href
  {https://doi.org/10.1103/PhysRevLett.127.268004} {\bibfield  {journal}
  {\bibinfo  {journal} {Phys. Rev. Lett.}\ }\textbf {\bibinfo {volume} {127}},\
  \bibinfo {pages} {268004} (\bibinfo {year} {2021})}\BibitemShut {NoStop}%
\bibitem [{\citenamefont {Talay}(1994)}]{talay1994numerical}%
  \BibitemOpen
  \bibfield  {author} {\bibinfo {author} {\bibfnamefont {D.}~\bibnamefont
  {Talay}},\ }\href@noop {} {\emph {\bibinfo {title} {Numerical solution of
  stochastic differential equations}}}\ (\bibinfo  {publisher} {Taylor \&
  Francis},\ \bibinfo {year} {1994})\BibitemShut {NoStop}%
\bibitem [{\citenamefont {Halperin}\ and\ \citenamefont
  {Hohenberg}(1969)}]{halperin1969scaling}%
  \BibitemOpen
  \bibfield  {author} {\bibinfo {author} {\bibfnamefont {B.}~\bibnamefont
  {Halperin}}\ and\ \bibinfo {author} {\bibfnamefont {P.}~\bibnamefont
  {Hohenberg}},\ }\href@noop {} {\bibfield  {journal} {\bibinfo  {journal}
  {Physical Review}\ }\textbf {\bibinfo {volume} {177}},\ \bibinfo {pages}
  {952} (\bibinfo {year} {1969})}\BibitemShut {NoStop}%
\bibitem [{\citenamefont {Gardiner}\ \emph {et~al.}(1985)\citenamefont
  {Gardiner} \emph {et~al.}}]{gardiner1985handbook}%
  \BibitemOpen
  \bibfield  {author} {\bibinfo {author} {\bibfnamefont {C.~W.}\ \bibnamefont
  {Gardiner}} \emph {et~al.},\ }\href@noop {} {\emph {\bibinfo {title}
  {Handbook of stochastic methods}}},\ Vol.~\bibinfo {volume} {3}\ (\bibinfo
  {publisher} {springer Berlin},\ \bibinfo {year} {1985})\BibitemShut {NoStop}%
\bibitem [{\citenamefont {Montagne}\ \emph {et~al.}(1996)\citenamefont
  {Montagne}, \citenamefont {Hern{\'a}ndez-Garc{\'\i}a},\ and\ \citenamefont
  {San~Miguel}}]{montagne1996numerical}%
  \BibitemOpen
  \bibfield  {author} {\bibinfo {author} {\bibfnamefont {R.}~\bibnamefont
  {Montagne}}, \bibinfo {author} {\bibfnamefont {E.}~\bibnamefont
  {Hern{\'a}ndez-Garc{\'\i}a}},\ and\ \bibinfo {author} {\bibfnamefont
  {M.}~\bibnamefont {San~Miguel}},\ }\href@noop {} {\bibfield  {journal}
  {\bibinfo  {journal} {Physica D: Nonlinear Phenomena}\ }\textbf {\bibinfo
  {volume} {96}},\ \bibinfo {pages} {47} (\bibinfo {year} {1996})}\BibitemShut
  {NoStop}%
\end{thebibliography}%

\end{document}